# Identification of parameters in building concentration dispersion model

**D. Calogine[*,1], H. Boyer[*], S. Ndoumbe**

[*]LPBS - Université de La Réunion
40 avenue de Soweto - 97410 Saint-Pierre, La Réunion

**Abstract**
The aim of this work is to simulate the pollutants transport in buildings. Focusing mainly on the presence of CO2, firstly we resolve the airflow equations for two typical validation cases, the Rao case and the IEA case. These numerical results are compared to the most known software and they are used to evaluate of the evolution of CO2 concentration in the different rooms. In order to obtain the different parameters and filters of the proposed model we use a statistical method based on Bayesian inference. The final comparison of results is coherent but a complementary experimental procedure is necessary to calibrate and refine the model.

## Introduction

Prediction of toxic substances presence in the living places becomes essential to preserve human health. The engineers of the building must be able to estimate, simulate and predict the concentration of substances defined as pollutants. The most frequently produced of them is the carbon dioxide ($CO_2$). People exhale $CO_2$ whenever they breathe and it is usually not toxic expect at very high concentration ($\geq 30{,}000$ ppm). We find $CO_2$ in nature everywhere and it is the major indication of atmospheric pollution. Parallel to the studies implemented for the storage of this substance in the ecological buildings, we must have a forecasting tool to alarm population of pollution's danger.
In order to add a module for the simulation of pollutant's transport into the CODYRUN software [4, 5], thermal building simulation software developed at LPBS (University of Reunion Island), we are interested in the resolution of the transport equations of concentration for multi-zones models. This concentration being driving by the circulation of air, we must simulate flow within buildings [1].
Several software packages are available for thermal simulations with airflow in buildings. Most frequently used are generally COMIS [20], CONTAM [17], EPS-r [11, 23], Airflow [7, 30], etc. There are various methods of numerical resolution. The nodal method [4, 5, 17, 11, 20] use macroscopic models, which represent room or entire building by a single node and calculate the flow through discrete paths such as doors, ducts, openings and cracks. Zonal method [29, 32, 33] split a room into different macroscopic homogeneous zones and characterizes the main driving flows in order to predict the spatial temperature distribution in the room. Computational Fluid Dynamics (CFD) tools [2, 7, 24] mostly known as microscopic modelling technique provide numerical solutions of the partial differential equations of governing airflow, contaminant, temperature and other related physical processes. CFD is mainly use for single room simulation or can also be combined with other previous methods.
For these different methods, the computing time generally grows according to the model complexity. The computing time, the setting up of the boundary conditions and other input parameters can be the drawbacks of CFD when practical studies are used.
For this work, we are interested in the first method giving usually satisfactory results. We are going to develop a model of transport of concentration in CODYRUN and to validate the proposed model we will successively use the COMIS software [8, 9, 10, 20, 21] and the CONTAM software [6, 10, 27] by solving two typical cases. The first case is the same as the one used by Haghighat and Rao, 1991). The second test case is called USER Test used in IEA Annex 23 [31].
The first part of this paper gives a brief introduction of the concentrations' equations to be considered. The recall of the method used to simulate airflow follows in second part. A brief presentation of the technique of resolution is given in order to introduce the comparison set up.
The third part focuses on a procedure of validation. The description of the two test cases, which we will simulate after their implementation in the cited software, is given.
The both sections following deal with numerical simulations of airflow and the first results for concentrations. These results obtained with the referenced software are confronted to our simulations. The last section is about to the adjustment of concentration transport. A Bayesian procedure is proposed to determine the parameters of our models.

[1] Corresponding author: didier.calogine@univ-reunion.fr

## 1. Transport equation of concentration

We are interested in the simulation of the dispersion of pollutant in buildings. We consider that buildings are composed of numbers of rooms (nodes) which are connected by openings to each other and to the outside. Ordinarily we consider equations based on the conservation of the concentration by taking into account their circulation in the different rooms. The balance equation is made locally using all interconnections of the building. The traditional equation of resolution of a dispersion problem of pollutant is summarized by the following relation [17, 22]:

$$V_i \frac{\partial C_i}{\partial t} = \sum_{j=1}^{N_j} Q_{j,i} C_j - \sum_{i=1}^{N_i} Q_{i,j} C_i + S_i \,, \qquad (1)$$

where $V_i$, $C_i$ and $S_i$ are the volume, the concentration and the generation of pollutant for the room $i$ respectively. The quantities $Q_{j,i}$ and $Q_{j,i}$ are the inflow and outflow of the room $i$ adjacent at the room $j$.
This equation given in a discretised form:

$$C_i^{n+1} = C_i^n + \Delta t * M^{-1}(DC_i + S_i) \qquad (2)$$

The concentration $C_i$ is then computed successively with computing time step $\Delta t$ and the evaluation to transition matrix $M^{-1}D$. The time step will be selected according to the order of magnitude of the flows calculated by the model. With the concentration of pollutant at time $t$, we will be able to evaluate its evolution knowing the transition matrix $M^{-1}D$. To do it we must determine the circulation of air in various building openings.

## 2. Simulation of air flow

We are going to represent one of the models used to simulate airflow in buildings. The chosen model is based on the work done by Haghighat and Rao (1991) [18,] and called system-theoretic approach. The derivation of this model uses matrix representations and his principle consists in calculating various pressures of each zone. The summarized principle is as follows: for a building having $N_Z$ zones and $N_0$ cracks (openings), $\Phi=\{\Phi_1, \Phi_2, \Phi_3, ..., \Phi_{N_z}\}$ and $Q = \{Q_1, Q_2, Q_3,..., Q_{N_O}\}$ are respectively vectors representing the pressure and airflow rates through openings. The pressure difference on both sides of these openings is given by the vector $\Delta P = \{\Delta P_1, \Delta P_2, \Delta P_3, ..., \Delta P_{N_O}\}$. The relation between pressures and flow at an opening depends on the aeraulic performance of the opening. If we use the model of small openings whose effective flow is governed by the following relation:

$$Q = f(\Delta P) = C\,(\Delta P)^n, \qquad (3)$$

where $C$ is the air's permeability coefficient in $m^3/(h \times Pa^n)$. Only the power law is implemented in this model version.

The key element in the Haghighat and Rao (1991) method is the matrix known as the "*incident matrix*" of the building interconnections. This matrix represents rooms' distribution in the building. It is defined by matrix elements denoted $\Pi_{ij}$ such that:

$$\underset{\substack{1\le i\le N_z\\ 1\le j\le N_O}}{\Pi_{ij}} = \begin{cases} 0 & \text{opening } j \text{ is not connected to zone } i \\ -1 & \text{for opening } j \text{ if flow comes out to zone } i. \\ 1 & \text{for opening } j \text{ if flow comes from zone } i \end{cases} \qquad (4)$$

So the pressure difference $\Delta P$ through all cracks or openings in the building is defined by the relation:

$$\Delta P = \Pi^T \Phi + P_f, \qquad (5)$$

where $P_f$ is called the "driving force". This force corresponds to the sum of all the aeraulic elements connected to the opening. The zones' pressures are determined as solution by using iterative Newton-Raphson method of the following matrix equation:

$$\Pi \cdot f(\Pi^T \Phi + P_f) = 0, \qquad (6)$$

where $F$ is the function defined by the equation (3) above. The knowledge of the vector $\Phi$, will allow us to deduce the pressure difference $\Delta P$ and consequently to obtain the required various flows $Q$.

Now with this formulation we can simulate airflow to estimate the distribution of pollutant in buildings. In order to justify our results obtained by this model we propose a comparison method.

## 3. Procedure of validation

### 3.1. *Presentation of validation software*

To allow us to compare our simulation results, we used two programs: COMIS and CONTAM. These software packages are used by a lot of people working on the thermal comfort of buildings [3, 10, 20, 21] and they require small computation. Both multizone airflow programs are going to allow us to adjust our model as a numerical experimentation method.

#### *3.1.1. COMIS software (Conjunction of Multi-zone Infiltration Specialists)*

The COMIS software results from the common working sessions (from October, 1998 to September, 1989) which took place at Lawrence Berkeley National Laboratory (LBNL). This software is now developed by the Scientific and Technical Centre for Building (CSTB) and the research institute “EMPA, Materials Science and Technology”. COMIS software simulates the airflow transfers and the transport of pollutants inside the buildings.
The COMIS’s library contains various active and passive components describing the building (openings, doors, windows, flows, regulators, ventilators) as well as other components defined by the relation between the flows and the apparent pressures. The mathematical foundations of COMIS are described by Fuestel (1999) and Lorenzzeti (2002).

#### *3.1.2. CONTAM software*

CONTAM is software of multi-zone model available to public. It is an indoor air quality and ventilation analysis computer program developed by the National Institute of Standards and Technology (NIST). These objectives are to determine the airflow and dispersion of pollutants in the buildings systems. Then it proposes to predict the exposure of occupants to contaminants. It uses a simplified zonal representation of buildings (Dols, 2001).
The model is based on the Axley’s method [41]. So airborne contaminants transport depends on the presence of all air movements (natural and mechanical) within the building system. For the physical phenomena's modelling, it uses the simplified relation of "well mixed zones", i.e. all the zone being taken into account by one node of calculation (Walton, 1989).

#### *3.1.3. CODYRUN software*

CODYRUN is a multi-zone software integrating ventilation and moisture transport transfer in buildings [Boyer, 96, 99]. It is developed by the “*Laboratoire de Physique du Bâtiment et des Systèmes”* (LPBS) at the university of Reunion island. In the software, a zone approach based on nodal analysis is employed. A coupled system describing thermal and airflow phenomena is resolved. In results, ambient thermal comfort is estimated for buildings systems.

We will use two tests cases for the models comparison. The first one is the same as the one used by Haghighat and Rao (1991). The second test case called USER Test used in IEA

Annex 23 [31]. For simplicity these two cases of study will be called in this paper as « *RAO* » and « *IAE* ».

3.2. *Presentation of test models*

*3.2.1. The RAO test case*

The *RAO* test case is a four-room building with eight airflow paths. The speed of the external wind is $5\ m \cdot s^{-1}$. It is taken to the maximum height of the building (roof). The outside and the neighbouring rooms through eight openings are represented in Fig. 1. The airflow paths are assumed to have zero humidity and have constant temperature.

*3.2.2. The IEA test case*

The *IEA* test case consists of a four-room building exposed to $2\ \mathrm{m \cdot s^{-1}}$ wind. As shown in Fig. 2, the building has ten airflow paths and outdoor temperatures higher than an indoor air temperature.

The implementation of the building geometry leads us to define the connections between the rooms and physical specificities of the openings such as wind pressure coefficients $C_p$, flow coefficient $C$, flow exponents $n$ and openings level $h$. All these parameter values of both test cases are given in Tab. 1 and Tab. 2.

We are looking for the various stationary mass flows through the openings for theses geometries.

**4. Numerical simulation of air flow**

We were interested in this part by the airflow through buildings. The use of the introduced method of part 3 leads us to define for RAO case and IEA case by the following matrices. Both linked in Figure 1 and Figure 2, based on Haghighat's definition, we introduce the "incident matrices":

$$\Pi_{\mathrm{RAO}} = \begin{pmatrix} 1 & 0 & 0 & 0 & -1 & -1 & 0 & 0 \\ 0 & 0 & 1 & 0 & 1 & 0 & -1 & 0 \\ 0 & 1 & 0 & 0 & 0 & 1 & 0 & -1 \\ 0 & 0 & 0 & 1 & 0 & 0 & 1 & 1 \end{pmatrix}, \qquad (7)$$

and

$$\Pi_{\mathrm{IEA}} = \begin{pmatrix} 1 & 0 & 0 & 0 & 0 & -1 & 0 & 0 & 1 & 0 \\ 0 & 1 & 0 & 0 & 0 & 0 & -1 & 0 & -1 & 1 \\ 0 & 0 & 1 & 0 & 0 & 0 & 0 & -1 & 0 & -1 \\ 0 & 0 & 0 & 1 & 1 & 1 & 1 & 1 & 0 & 0 \end{pmatrix}. \qquad (8)$$

*4.1.1. RAO case*

All results obtained with the software CONTAM, COMIS and the CODYRUN software are represented on the buildings below. Flow's directions are represented too. We specified the numerical values given in the reference [25] corresponding to computed values by Tuomaala.

*4.1.2. IAE case*

We note results of airflow in the same order of magnitude in general. We note small differences in results between different software and the developed model. This great geometric difference of this one in comparison with the previous case is the presence of a transverse common room including five openings.

## 5. Simulation of concentration dispersion

### 5.1. *Model of dispersion*

Having the velocity fields and flows crossing each room, we are interested now by simulation of dispersion of pollutants in the two standard buildings [13, 28].
Referring to part 1 and figures 1 and 2, these equations are written as follow. If $C_i$ denote the present concentration in the room *i*, $V_i$ his volume and *m* the mass flows, calculated before, by taking into account the difference between inside and outside, the mass conservation equation for these concentrations in the different rooms is written as:

$$V_i \frac{\partial C_i}{\partial t} = \sum_{j,k} m_j C_k - \sum_{l,m} m_l C_m + S_i . \tag{9}$$

Applying this relation for the IEA model, we obtain the following system of equations for concentrations $C_1$, $C_2$, $C_3$ and $C_4$:

$$\begin{aligned}
V_1 \frac{\partial C_1}{\partial t} &= m_1 C_0 - m_6 C_1 - m_9 C_1 + S_1 \\
V_2 \frac{\partial C_2}{\partial t} &= m_2 C_0 - m_{10} C_2 + m_9 C_1 - m_7 C_2 + S_2 \\
V_3 \frac{\partial C_3}{\partial t} &= m_{10} C_2 + m_8 C_4 - m_3 C_3 + S_3 \\
V_4 \frac{\partial C_4}{\partial t} &= m_6 C_1 + m_7 C_2 - m_8 C_4 - m_4 C_4 + m_5 C_0 + S_4
\end{aligned} \tag{10}$$

The same procedure applied to the RAO model, we obtain the analogue following system of equations:

$$V_1 \frac{\partial C_1}{\partial t} = m_2 C_0 + m_6 C_3 - m_5 C_2 + S_1$$
$$V_2 \frac{\partial C_2}{\partial t} = m_5 C_1 + m_7 C_4 - m_3 C_2 + S_2$$
$$V_3 \frac{\partial C_3}{\partial t} = m_1 C_0 - m_6 C_3 - m_8 C_3 + S_3 \qquad (11)$$
$$V_4 \frac{\partial C_4}{\partial t} = m_8 C_3 - m_4 C_4 - m_7 C_4 + S_4$$

At this level, the procedure of comparison consists in confronting the results of simulation acquired with software CONTAM and those of the first-order model of equations (10) and (11).
To do it we define a test case by imposing an initial concentration in one of the rooms of a building. Then in order to obtain a comparison of our results with CONTAM software, we simulated with this software the same geometry and the same initial conditions of $C0_2$ concentration. Of course we can use independently the RAO or the IEA case at our disposal.

5.2. *Introduction of displacement factor and filters*

For the IEA case, we will be interested in the variation of the concentration of $C0_2$ in the different rooms of the building. We considered for this simulation an initial concentration of 100 kg/m$^3$ of pollutant in a room located at the ground floor, the room number 3 of Figure 1.

In our model, we also initiated the concentration with 100 kg/m$^3$ in the room at the ground floor. So we have to cancel the various sources of pollutants: $S_1 = S_2 = S_3 = S_4 = 0$ and we specified the initial concentrations $C_3^0 = 100$ kg m$^{-3}$ and $C_1^0 = C_2^0 = C_4^0 = 0$ kg m$^{-3}$. We obtain the time evolutions of concentrations in the rooms of Figure 10.
These results seem coherent with those of the previous figure. We notice the reduction of initial concentration, accumulations and then discharges in the other rooms according to time. This kind of classical evolution is also noticed by CONTAM software. However a confrontation of the two results reveal a difference for the speed of accumulation and the reduction of curves. This great difference is due on the one hand to the little disparity in the calculation of flows (cf. results part 4) and on the other hand to the difference in dispersion model used. Therefore we must look closely at the model.
Indeed, the model used by CONTAM software [6] can be summarized by the following quantities.
If we denote $m_{\alpha,i}$ defined by notation $m_i.C_{\alpha,i}$, the product of the mass of contaminant α in zone $i$, with $m_i$ the mass of air in the zone $i$ and $C_{\alpha,i}$ the mass fraction of concentration α, the time variation of this mass of contaminant is written as:

$$\frac{dm_{\alpha,i}}{dt} = -\underbrace{R_{\alpha,i} C_{\alpha,i}}_{displacement} - \underbrace{\sum_j F_{i,j} C_{\alpha,i}}_{\text{Airflow outside the zone}} + \underbrace{\sum_j F_{j,i}(1-\eta_{\alpha,j,i}) C_{\alpha,i}}_{\text{Airflow inside the zone } (\eta \text{ filter})} + \underbrace{m_i \sum_\beta \kappa_{\alpha,\beta} C_{\beta,i}}_{chemical\ reaction\ 1^{st}\ order} + \underbrace{G_{\alpha,i}}_{source\ (generation)}, \qquad (12)$$

where $F_{i,j}$ is the mass flow between rooms of indices *i* and *j*, $G_{\alpha,i}$ is the term source, $\kappa_{\alpha,\beta}$ are stœchiometric parameters linked to the chemical reactions, $R_{\alpha,i}$ is the term of displacement and $\eta_{\alpha,i,j}$ are the filter parameters.

This is a conservation equation taking into account all the contributions intervening within the room. The terms similar to our model are those of airflow between the room and the term of generation (source). The modelling of the equation (12) uses filters at openings and an additional term of displacement.

According to previous models results, the displacement term turns out not to be negligible for the estimation of concentration transfer within the building. We are going to propose a procedure of Bayesian method to identify these parameters.

## 6. Bayesian estimation of parameter

### 6.1. *Introduction to Bayesian method*

To solve this difference in numerical results we are looking for a supplementary displacement term in the form $\alpha\, C^{\beta}$ and the filter parameters for our model.

The technique employed looks like an optimisation method (fitting procedure). For each room, the parameters $\alpha$, $\beta$ and $\psi_{i,j}$ associated concentration will be determined by Bayesian identification.

Bayesian probability theory is currently experiencing an increase in popularity in the sciences as a means of probabilistic inference [19]. Bayesian methods can be used either for model selection problems or for parameter estimation problems. Advantages of Bayesian approach include the aptitude to incorporate prior information about model parameters into the analysis. This leads to obtain inference (up to Monte Carlo error) without need for large sampling approximations and notably, the development of the WinBUGS software [37, 38] gave an overview of using WinBUGS for Bayesian modelling and reviewed several models to estimate the sensitivity and specificity of multiple diagnostic tests. A good review of the Bayesian approach is given by Jaynes and Loredo [12, 16].

For parameter estimation problems, Bayesian inference deals with the estimation of the values of *p* model parameters $\theta = (\theta_1, \theta_2, \cdots, \theta_p)$ about which there may be some prior beliefs. These prior beliefs can be expressed as a probability density function (p.d.f) called *prior*, $\pi(\theta)$ and may be interpreted as the probability placed on all possible parameter values before collecting any new data. The dependence of observations (or measurements) $D = (d_1, d_2, \cdots, d_N)$ on the *p* parameters $\theta$ can be also expressed as a p.d.f.: $L(D/\theta)$ called the *likelihood function*. The latter is used to update the prior beliefs on $\theta$, to account for the new data *D*. This updating is done through the Bayes' theorem:

$$\pi(\theta / D) = \frac{\pi(\theta) L(D/\theta)}{\int_{\theta} \pi(\theta) L(D/\theta) d\theta} \tag{13}$$

Where $\pi(\theta / D)$ represents the *posterior* p.d.f and expresses the values of the parameter after observing the new data. In other words, the *prior* is modified by the *likelihood* function to yield the *posterior*.
In the Bayesian framework, uncertainties in parameter values are naturally assessed. For instance, the position of the maximum of the *posterior* p.d.f represents a best estimate of the parameter; its width or spread about this optimal value gives an indication of the uncertainty in the estimate of the parameters.

In this paper, in order to estimate the parameters of our models by a Bayesian approach, the simulations are made with "WinBUGS". The R [39] package R2WinBUGS [40] was used to call WINBUGS and export results in R. As we have no knowledge about the parameter values before collecting new data, we attributed a uniform distribution for the prior.

6.2. *Comparison of results*

Firstly the procedure consists in identify the filter parameters $\psi_{j,k}$ associated at each input flux referenced $j$ for the different rooms of index $k$. Secondly we have to take into account a displacement term depending of concentration. The masse conservation equation (9) can be rewritten as:

$$V_{\mathrm{i}} \frac{\partial C_i}{\partial t} = -\alpha \cdot C_i^{\beta} + \sum_{j,k} \psi_{j,k} m_j C_k - \sum_{l,m} m_l C_m + S_i , \qquad (14)$$

where index *i* refers to the room considered, index *j* and *k* are associated to the inflow and indexes *l* and *m* to the outflow.
We are looking for the parameters denoted $\alpha$, $\beta$ and $\psi_{j,k}$ which appear in equation (15).
The procedure is implemented using the BRugs package [15].
To determine the model's parameters, we are interested by the evaluation of probability densities provided by the software. In order to determine the appropriate parameters which satisfy the model at best, these values are given by the maximum of the probability densities. With the initial values $C_3^0 = 100$ kg m$^{-3}$ and $C_1^0 = C_2^0 = C_4^0 = 0$ kg m$^{-3}$, for the rooms A, B, C and D to initialize our model, we statistically estimate the parameters which adjust the CONTAM software results. These reference simulations are done with a time step of one second.
By this method, the maximum of probability density curve given by the BRugs package will nearly predict the best candidates.
The obtained values are homogeneous for the four rooms. Concerning the replacement terms, the amplitude terms are about $10^{-4}$ and the exponential terms is lower than one. These global terms are moderated in amplitude: so we deduce than the displacement of concentration is mainly drived by the airflow.
For the filter terms, the results give a majority of factor lower than one. These $\psi$ factors correspond to the parameters denoted $(1-\eta)$ in CONTAM software. The attenuation of the filters introduced by the $\eta$ factor is well effective in this kind of simulation. Otherwise we obtain an estimation of 1.207 for the parameter $\psi_{(7,4)}$ even if the airflow calculation for this crack is satisfactory.
Afterward the selection of these parameters allows us to compare our adjusted model with the concentration of computed by CONTAM software as in the preceding case. In order to do that, we plot the evolution of contaminant concentration in every room. The decrease of initial concentration in the room C is observed and the model matches accurately with the reference data. The concentration increasing is represented in the three other rooms at first followed by the decreasing phase. These simulations are in total accordance to the software. The assumed overestimation of the parameter $\psi_{(7,4)}$ seems to have a limited influence on the concentrations. At this stage, we can assert that the RAO model described by eq. 15 is suitable.

We apply the same estimation procedure for the IEA model. According to the circulation of airflow in the building we choose to introduce the initial values $C_1^0 = 100$ kg m$^{-3}$, $C_2^0 = 75$ kg m$^{-3}$, $C_3^0 = 50$ kg m$^{-3}$ and $C_4^0 = 0$ kg m$^{-3}$.
The estimation of the smoothed kernel density for each parameter's identification are quantified and reported. As for the preceding case most of parameters are coherent between them. Otherwise the $\alpha$ value of the room D ($\approx 10^{-2}$) seems to be too wide compared to the others ($\approx 10^{-4}$) and for the room C the parameter $\psi_{(10,2)}$ obtained is superior to one.
Using identified parameters, we can compare the new results for the IEA case to those simulated by the CONTAM software. The evolution of concentration for each room is then evaluated.
We note a little difference in the decreasing form of the concentrations for the rooms A, B and C. These amplitudes disparity is significant but acceptable. Otherwise the difference is accentuated for the room D. Our model underestimates the peak of concentration at this place. This failing is due to the strong value of $\alpha$, responsible to the discharge phase and the weakness to parameters $\psi_{(6,1)}$ and $\psi_{(7,2)}$. These filter parameters seem to be close to 1. When one parameter is too small, a second one needs to compensate for the adjustment like for the room C in the IEA case and for the room B in the ROA case. Unfortunately we have not obtained convergence with the BRugs package for this scale of values.

6.3. *Remark*

The procedure proposed in the previous paragraph not allows us to obtain all the parameter of the model (12). In fact the filters parameters associated to openings with a positive inflow are cancelled by the null exterior concentration. We need to specify several different outdoor concentrations in CONTAM software, one concentration for each opening, in order to validate our model. A complementary experimental procedure of validation seems to be necessary to adjust some specific parameters and to set up a "crossed estimate" taking into account several groups of parameters. For example, an experimental procedure of combined opened/closed crack will be necessary to characterise one after another all the openings.

## 7. Conclusion

These simulations enabled us to test two traditional cases, the RAO case and the IEA case with COMIS software and CONTAM software. The results are summarized on Fig. 7 and Fig. 8 for airflow simulations.
In order to implement a contaminant dispersion model in CODYRUN software we presented a module to calculate in a first step, the flows at the openings and in the second step, the estimation of time variation for one or several concentrations in the different rooms (in absence of chemical reactions).

This model is simple and effective. It gives values comparable to those of others commercial software of reference. The values of concentrations are underestimated when we neglect filter at the openings. These filters tend to attenuate the exchange between the rooms.

The simulations have showed the relative importance of displacement term in the formulation of contaminant variation.

A first approach of parameters estimation is proposed. The results obtained using Bayesian inference are similar to those obtained with commercial software.

The following stage will consist in taking into account the thermal equations. Indeed, the temperature field may modify the airflow. Consequently, the distribution of concentration could be significantly affected.

Furthermore, we propose to set up a procedure of validation thanks to the use of concentration sensors in order to dispose experimental measurements [14]. These data will allow us to complete the model validation procedure and if necessary to refine the model of pollutants' dispersion.

Otherwise a procedure of model comparison is possible to refine the model of displacement. It will be supplemented by the correlation of searched parameters with the physical and geometric properties of openings to offer a practical tabulation of parameters $\alpha$ and $\beta$.

## Acknowledgments

This work was supported by the "Region Reunion" within the research program 2006-2009 in partnership with the University of Reunion Island.

## Reference

[1] Allard Francis, Natural ventilation in buildings, A design handbook, *James and James (Science Publishers) LTD*, London, 1998.

[2] Bastide A., Lauret P., Garde F., Boyer H. Building energy efficiency and thermal comfort in tropical climates: Presentation of a numerical approach for predicting the percentage of well-ventilated living spaces in buildings using natural ventilation, *Energy and Buildings*, 38, 9, pp. 1093-1103, 2006.

[3] Bossaer A., Ducarme D., Wouters P., Vandaele L., An example of model evaluation by experimental comparison: pollutant spread in a apartment, *Energy and Building*, vol. 30, pp. 53-59, 1999.

[4] Boyer H., A. P. Lauret, L. Adelard and T. A. Mara, Building ventilation: a pressure airflow model computer generation and elements of validation *Energy and Buildings*, Volume 29, Issue 3, January 1999, Pages 283-292.

[5] Boyer H., J. P. Chabriat, B. Grondin-Perez, C. Tourrand and J. Brau. Thermal building simulation and computer generation of nodal models *Building and Environment*, Volume 31, Issue 3, May 1996, Pages 207-214.

[6] Dols Start, A tool for modelling airflow & contaminant transport, *ASHRAE Journal*, Practical guide.2001.

[7] Evola G., Popov V., Computational analysis of wind driven natural ventilation in buildings, Energy and Buildings, 38, pp. 491-501, 2006.

[8] Feustel H.E., COMIS- an international multi-zone air-flow and contaminant transport model, *Energy and Building*, vol. 30, pp. 3-18, 1999.

[9] Haas Anne, Andreas Weber, Viktor Dorer, Werner Keilholz and Roger Pelletret, COMIS v3.1 simulation environment for multizone air flow and pollutant transport modelling, *Energy and Buildings*, Volume 34, Issue 9, October 2002, Pages 873-882.

[10] Haghighat F., Rao J., A comprehensive validation of two airflow models – COMIS and CONTAM. Indoor air, 6, pp. 278-288, 1996.

[11] Heim Dariusz, Clarke Joe A., Numerical modelling and the thermal simulation of PCM-gypsum composites with ESP-r, Energy and Building, 36, pp. 795-805, 2004.

[12] Jaynes E.T. Bayesian methods: an introductory tutorial. Maximum entropy and bayesian methods in applied statistics, *Cambridge University Press*, Cambridge, UK, 1986.

[13] Klobut K., P. Tuomaala, K. Siren, O. Seppanen, Simulation calculation of airflows, temperatures and contaminant concentration in multi-zone buildings, Air movement & ventilation control within buildings, *12$^{th}$ AIVC Conference*, Ottawa, Canada, 1991.

[14] Kozdron-Zabiegada Bozenia, Jacek Namiesnik, Andrzej Przyjazny, Use of passive dosimeters for evaluation of the quality of indoor and outdoor air, *Indoor Environ*, vol. 4, pp. 189-203, 1995.

[15] Ligges U. The BRugs package. OpenBRugs documentation. 2006.

[16] Loredo T.J. From Laplace to supernova 19871: Bayesian inference in astrophysics. Maximum entropy and bayesian methods in applied statistics, *Cambridge University Press*, Cambridge, UK, 1990.

[17] Matsumoto H., Hayashi S., A multi-zone ventilation model with contaminant emissions from building materials, AIVC 12045, *Proceedings of Roomvent 98: 6th International Conference on Air Distribution in Rooms*, Volume 1, pp. 5, 1998.

[18] Haghighat F., Rao J., Computer-aided building ventilation system design – a system-theoretic, *Energy and buildings*, vol. 17, pp. 147-155, 1991.

[19] Riviere C., Lauret P., Manicom Ramsamy J.F., Page Y. A Bayesian Neural Network approach to estimating the Energy Equivalent Speed. *Accident Analysis & Prevention*, 38, 2, Pages 248-259, 2006.

[20] Rodriguez E.A., Allard F., Coupling COMIS airflow model with other transfer phenomena, *Energy and buildings*, 18, 147-157, 1992.

[21] Roulet Claude-Alain, Furbringer Jean-Marie, Cretton Pascal, The influence of the user on the results of multizone air flow simulations with COMIS, *Energy and Buildings*, vol. 30, pp. 73-86, 1999.

[22] Sohn M.D., Small M.J., Parameter estimation of unknown air exchange rates and effective mixing volumes from tracers gas measurements for complex multi-zone indoor air models, *Building and Environment*, vol. 34, pp. 293-303, 1999.

[23] Strachan P.A., Kokogiannakis G., Macdonald I.A., History and development of validation with the ESP-r simulation program, *Building and Environment*, 2006.

[24] Tan Gang, Glicksman Leon R., Application of integrating multi-zone model with CFD simulation to natural ventilation prediction, *Energy and building*, vol. 37, pp. 1049-1057, 2005.

[25] Tuomaala P., New building air flow simulation model: Theoretical basis, Building Serv. Eng. Res. *Technol.*, 14(4) 151-157 (1993)

[26] Tuomaala P., Rahola J., Combined Air flow and Thermal simulation of buildings, *Building and Environment*, vol. 30, n° 2, pp. 255-265, 1995.

[27] Walton George N., CONTAM96 User Manual. Report NISTIR 6056, U.S. Department of Commerce, National Institute of Standards and Technology, Gaithersburg, 1997.

[28] Water J.R., M.W. Simons, The evaluation of contaminant concentration and Air flows in a multi-zone model of a building, *Building and Environment*, vol. 22, n°4, pp. 305-315, 1987.

[29] Wurtz E., Mora L., Inard C. An equation-based simulation environment to investigate fast building simulation, Building and Environment, 41, pp. 1571-1583, 2006.

[30] Yuan Feng-Dong, You Shi-Jun, CFD simulation and optimisation of the ventilation for subway side-platform, Tunnelling and Underground Space Technology, 22, pp. 474-482, 2007.

[31] Furbringer J.-M., Roulet C.-A, Borchiellini, Annex 23: Multizone Air Floz Modelling, Evaluation of Comis, LESO-EPFl, 1996.

[32] Musy M., Wurtz E., Sergent A., Buildings air-flow simulations:automatically generated zonal models, *7$^{th}$ Internationa IBPSA Conference, Riode Jainero, Brazil*, 1, 593-600, 2001

[33] Haghighat F., Li Y., Megri A. C., Development and validation of a zoanl model – POMA, *Buildings and Environment*, 36, pp 1039-1047, 2001.

[34] Megri A. C., Snyder M., Musy M., Building zonal thermal and airflow modelling – A review, *International. Journal of Ventilation*, 4 (2), pp. 177-188

[35] Li Y., Delsante A., Sysmons J., Prediction of natural ventilation in buildings with large openings, *Buildi ngs and Environment*,35(3), pp. 191-206, 2000

[36] Lorenzetti, 2002 D.M. Lorenzetti, Computational aspects of nodal multizone airflow systems, Building and Environment 37 (2002), pp. 1083-1090.

[37] Spiegelhalter D., Thomas A., Best N., Gilks W., BUGS: Bayesian inference using Gibbs sampling version 0.50. MRC Biostatistics Unit, Cambridge, 1996

[38] Spiegelhalter D., Thomas A., Best N., Lunn D., WinBUGS user manual-version 1.4. MRC Biostatistics Unit, Cambridge, UK.

[39] Ihakaa R., Gentleman R., R: a language for data analysis and graphics. *J. Comput. Graph. Stat.* 5. pp. 299-314, 1996

[40] Sturtz S., Ligges U., Gelman A., R2WinBUGS: a package for running WinBUGS from R. J. Statist. Software 12, 1-16, 2005.

[41] Axley, J.W. 1988. "Progress Toward a General Analytical Method for Predicting Indoor Air Pollution in Buildings, Indoor Air Quality Modeling Phase III Report", NBSIR 88-3814, National Bureau of Standards (U.S.).

[42] Walton, G.N. 1989. *AIRNET - A Computer Program for Building Airflow Network Modeling*. NISTIR 89-4072. National Institute of Standards and Technology.

## Tables

| Path | $C_p$ | C [$m^3s^{-1}Pa^{-n}$] | n | Height [m] | Path | $C_p$ | C [$m^3s^{-1}Pa^{-n}$] | n | Height [m] |
|---|---|---|---|---|---|---|---|---|---|
| 1 | +0.9 | 0.005 | 0.66 | 5.0 | 5 | 0.0 | 0.015 | 0.66 | 5.5 |
| 2 | +0.3 | 0.008 | 0.66 | 1.0 | 6 | 0.0 | 0.020 | 0.66 | 3.0 |
| 3 | -0.4 | 0.007 | 0.66 | 5.0 | 7 | 0.0 | 0.020 | 0.66 | 3.0 |
| 4 | -0.3 | 0.009 | 0.66 | 2.0 | 8 | 0.0 | 0.015 | 0.66 | 1.5 |

Tab. 1. Flow element parameters for the *RAO* test case

| Path | $C_p$ | C [$m^3s^{-1}Pa^{-n}$] | n | Height [m] | Path | $C_p$ | C [$m^3s^{-1}Pa^{-n}$] | n | Height [m] |
|---|---|---|---|---|---|---|---|---|---|
| 1 | +0.2 | 0.02 | 0.65 | 2.0 | 6 | 0.0 | 2.0 | 0.50 | 1.0 |
| 2 | +0.4 | 0.02 | 0.65 | 5.0 | 7 | 0.0 | 0.04 | 0.50 | 4.0 |
| 3 | +0.5 | 0.02 | 0.65 | 8.0 | 8 | 0.0 | 0.04 | 0.50 | 7.0 |
| 4 | -0.4 | 0.02 | 0.65 | 9.0 | 9 | 0.0 | 0.004 | 0.50 | 3.0 |
| 5 | -0.3 | 0.02 | 0.65 | 1.0 | 10 | 0.0 | 0.004 | 0.50 | 6.0 |

Tab. 2. Flow element parameters for the *IEA* test case

| **Room** | $\alpha$ | $\beta$ | $\psi_{(j,k)}$ |
|---|---|---|---|
| **A** | -0.0002808 | 0.395 | $\psi_{(6,3)} = 0.9882$ |
| **B** | -0.000964 | 0.8738 | $\psi_{(5,1)}$ : 0.2873<br>$\psi_{(7,4)}$ : 1.207 |
| **C** | -0.0007427 | 0.6436 | |
| **D** | -0.0009862 | 0.219 | $\psi_{(8,3)}$ : 0.992 |

Tab. 3. Identification of parameters for RAO model

| **room** | $\alpha$ | $\beta$ | $\psi_{(j,k)}$ |
|---|---|---|---|
| **A** | 0.003331 | 0.9971 | |
| **B** | 0.001241 | 0.9593 | $\psi_{(9,2)}$ : 0.7695 |
| **C** | 0.0003706 | 0.3923 | $\psi_{(10,2)}$ : 1.286<br>$\psi_{(8,4)}$ : 0.565 |
| **D** | 0.01157 | 0.9866 | $\psi_{(6,1)}$ : 0.5404<br>$\psi_{(7,2)}$ : 0.5441 |

Tab. 4. Identification of parameter for model IEA

**Figures**

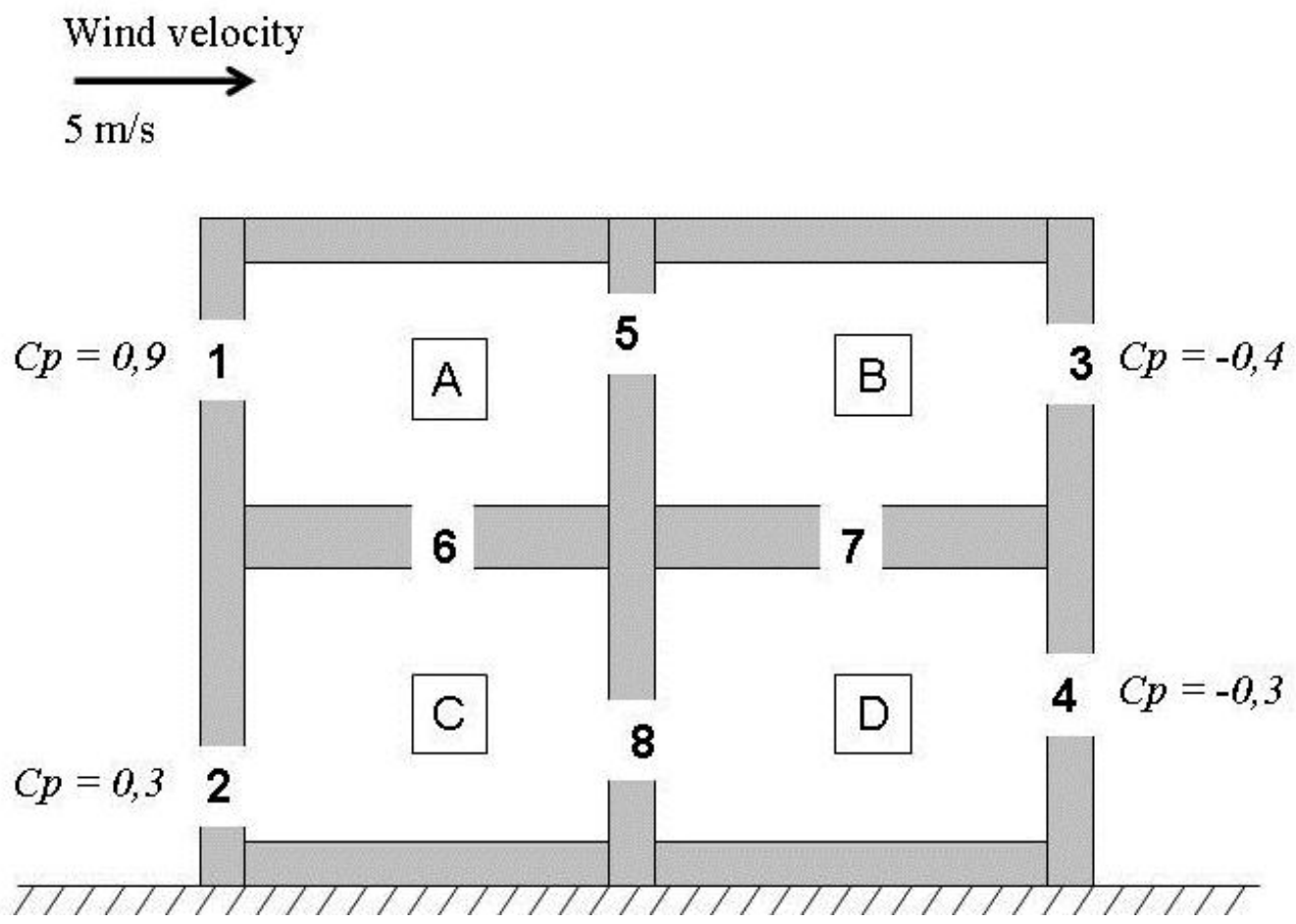

Fig. 1. The « RAO » building

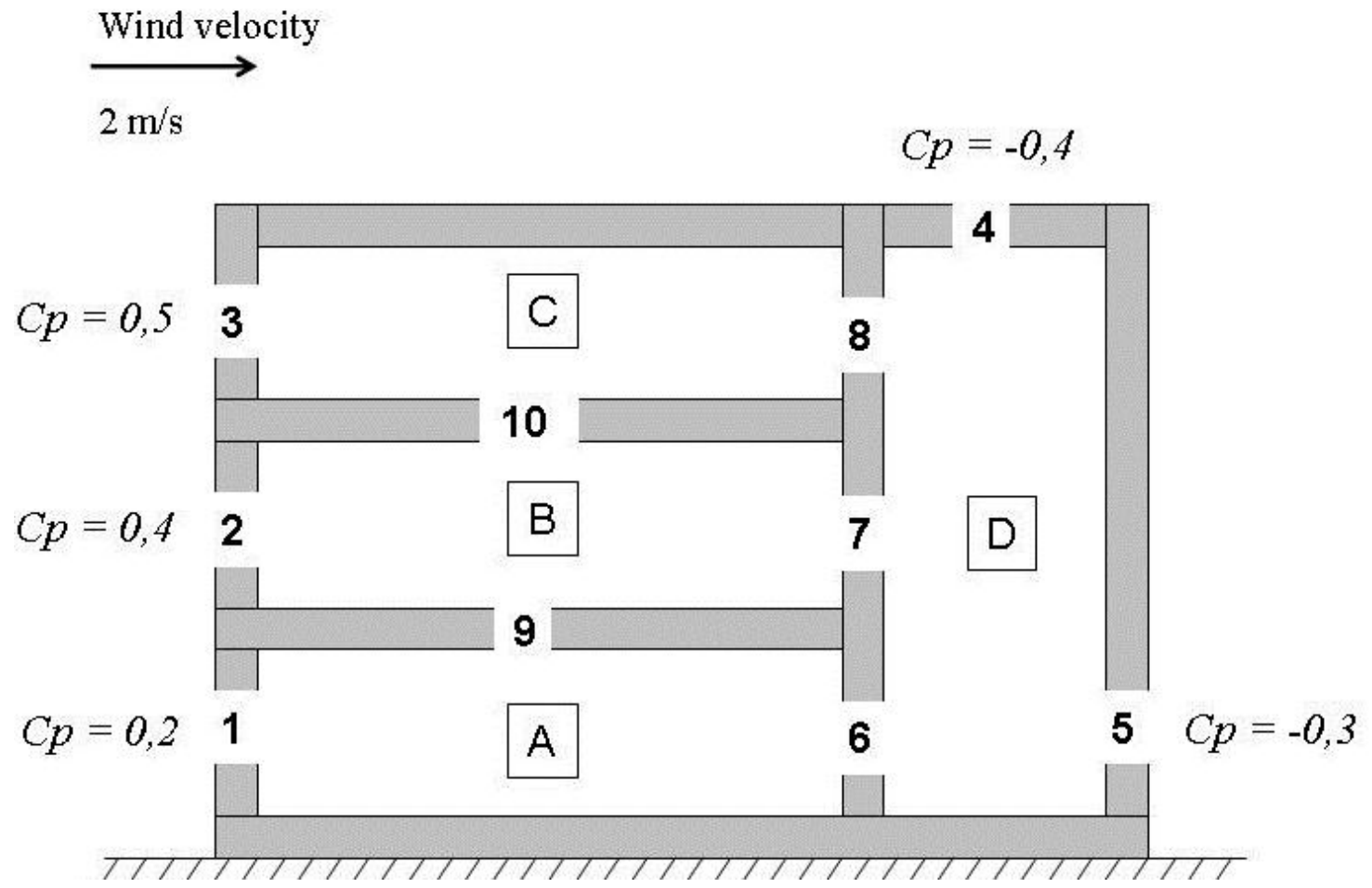

Fig. 2. The « IEA » building

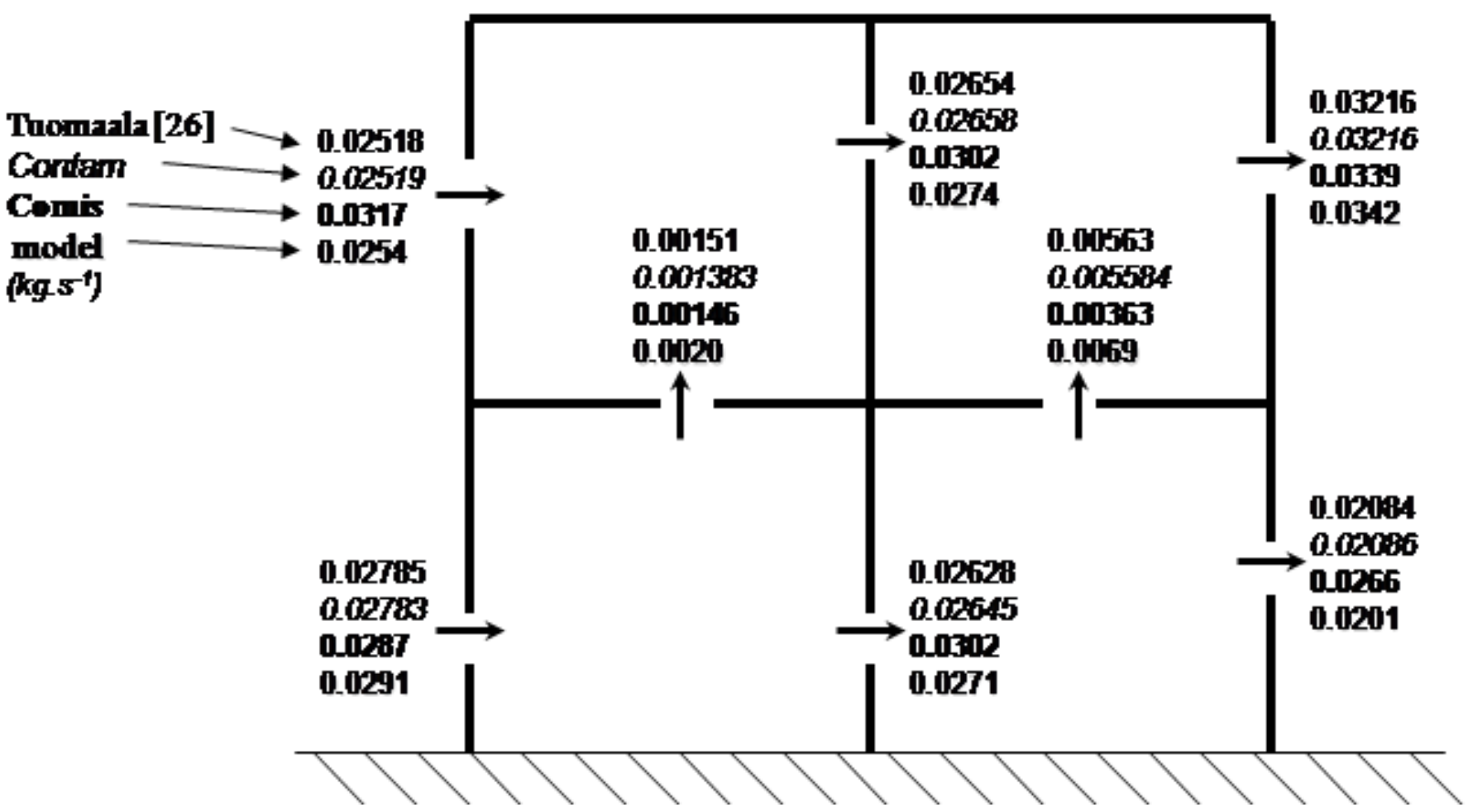

Fig. 3. Results for RAO building

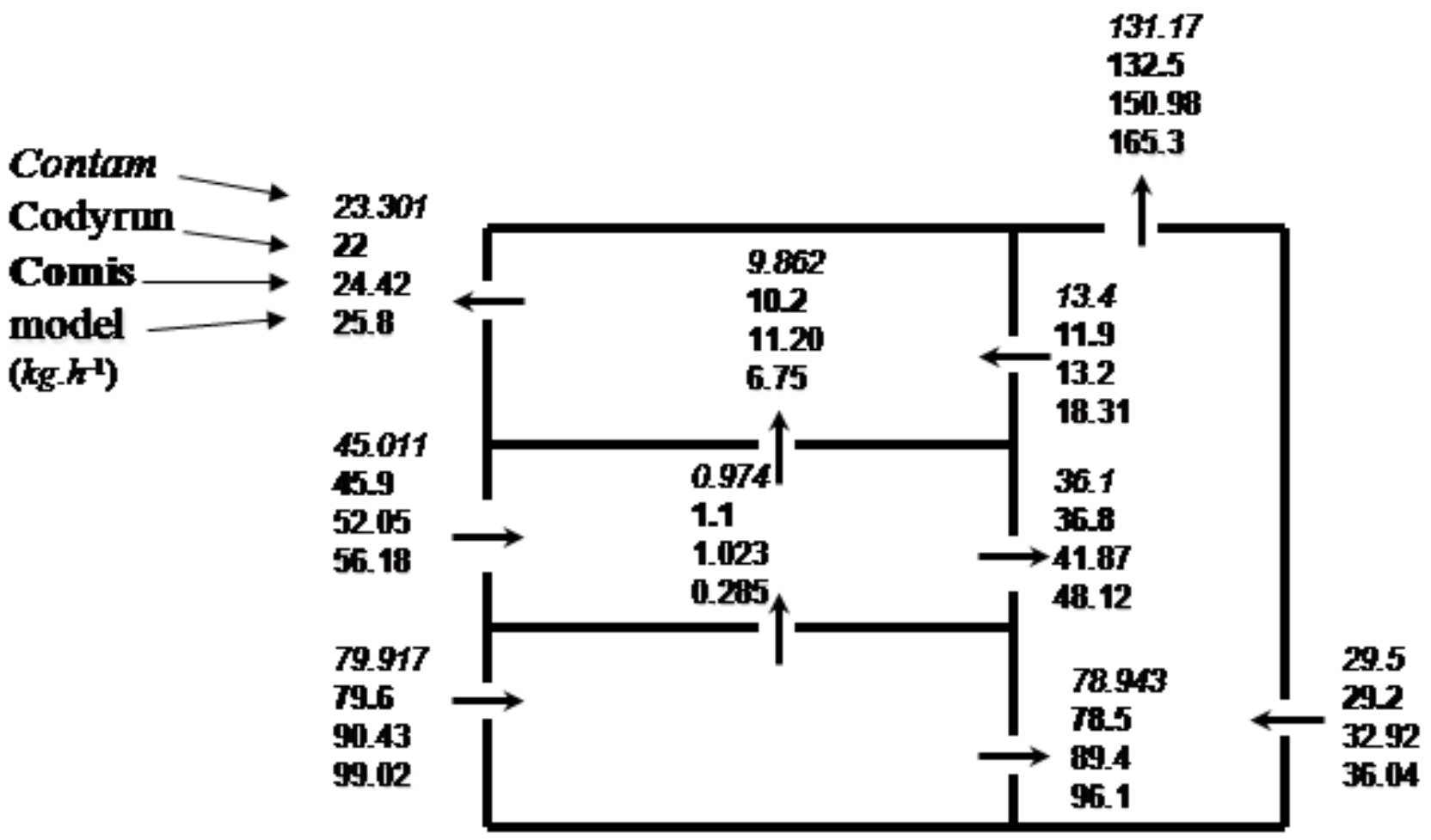

Fig. 4. Results for RAO building

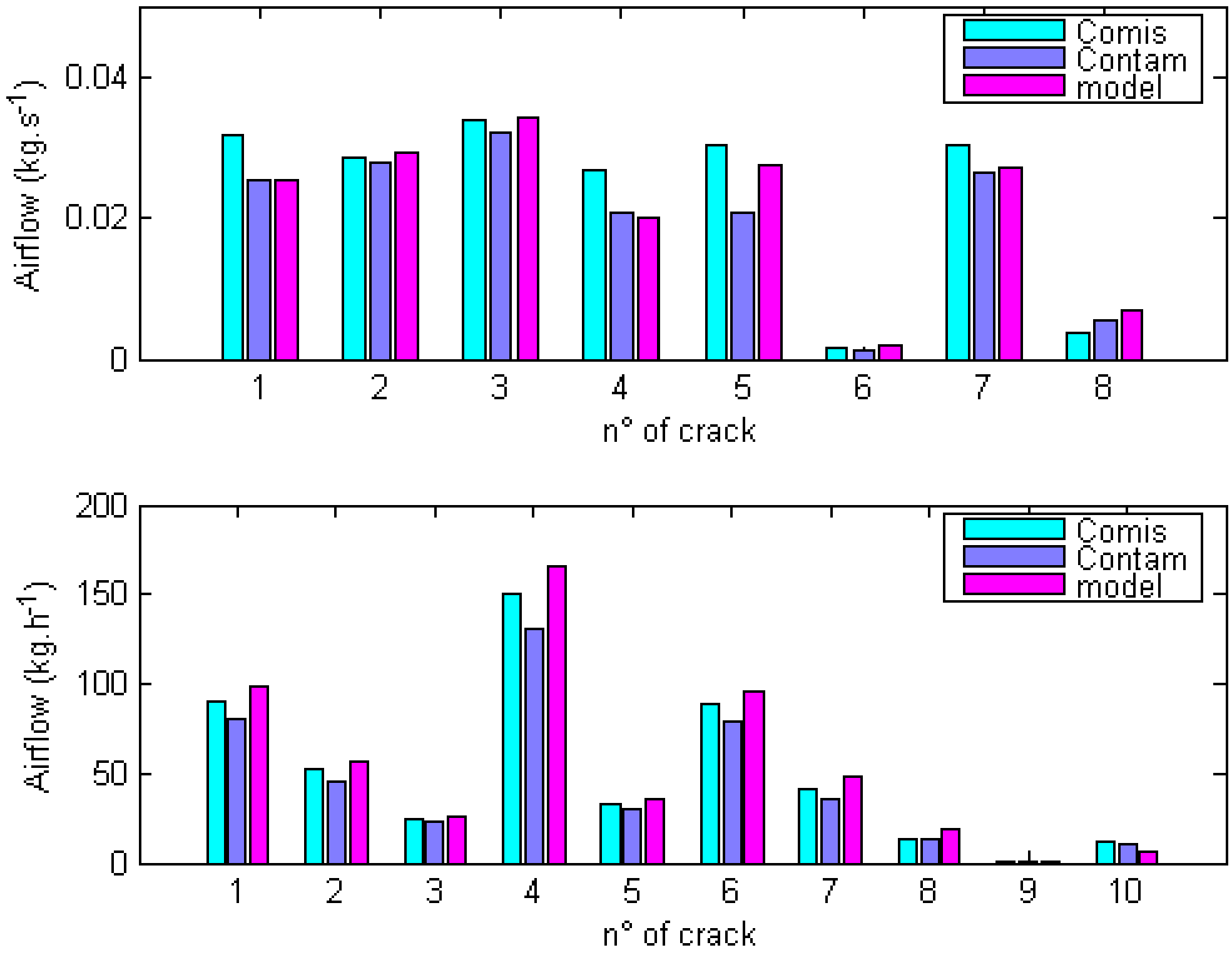

Fig. 5. Airflow for each crack (RAO case above and IEA one down)

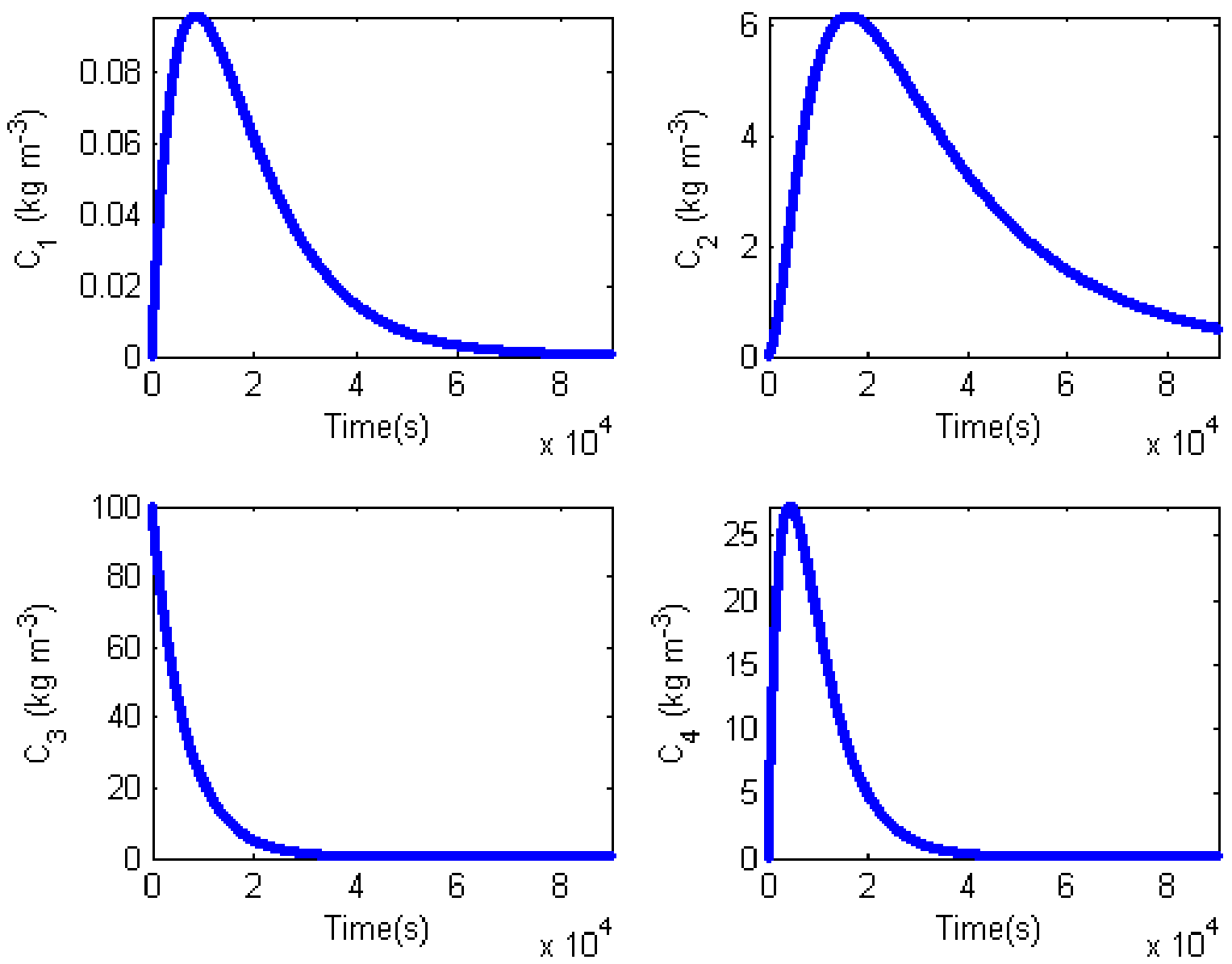

Fig. 6. Variation of $C0_2$ in rooms.

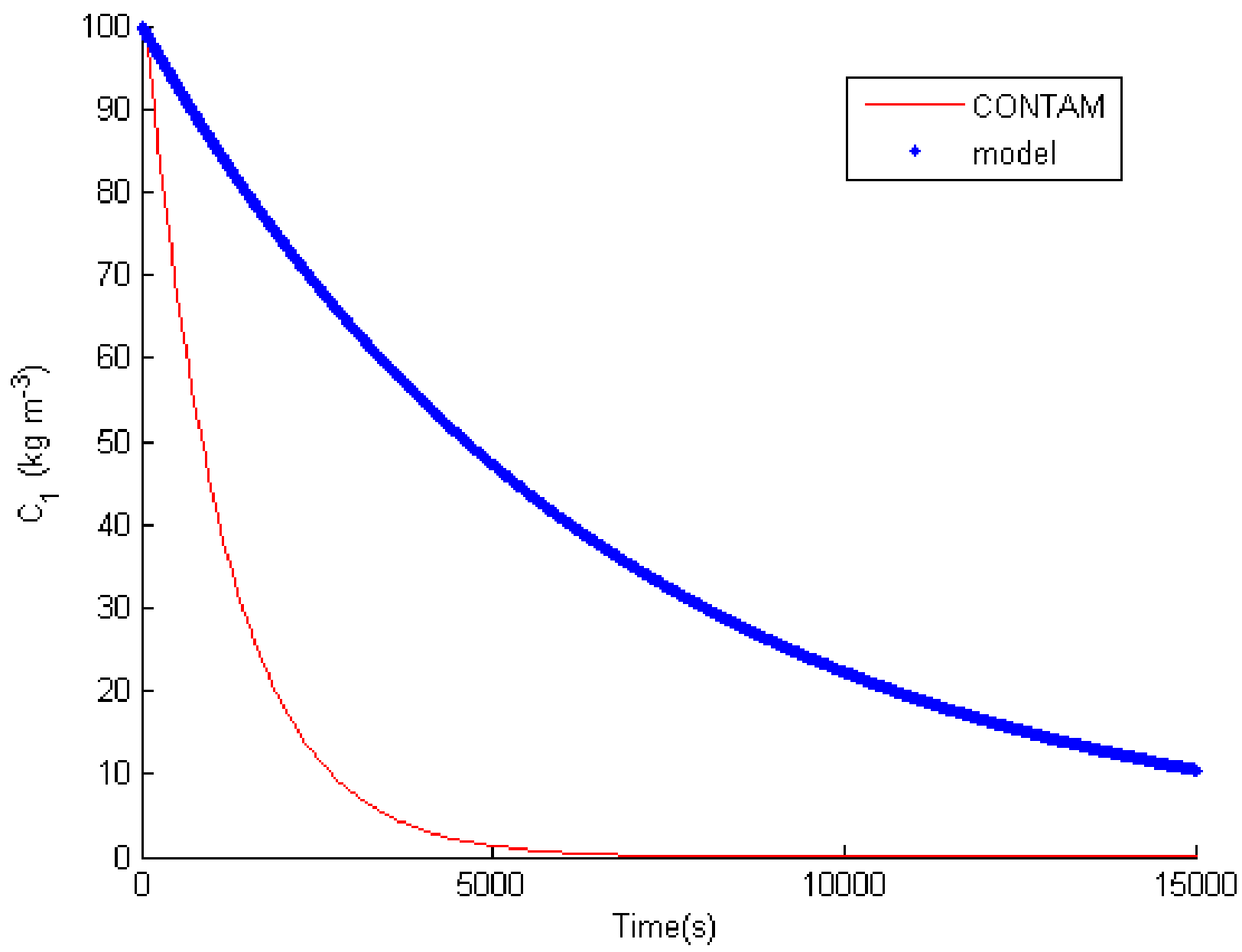

Fig. 7. Variation of $C0_2$ concentration in the same room (IEA case)

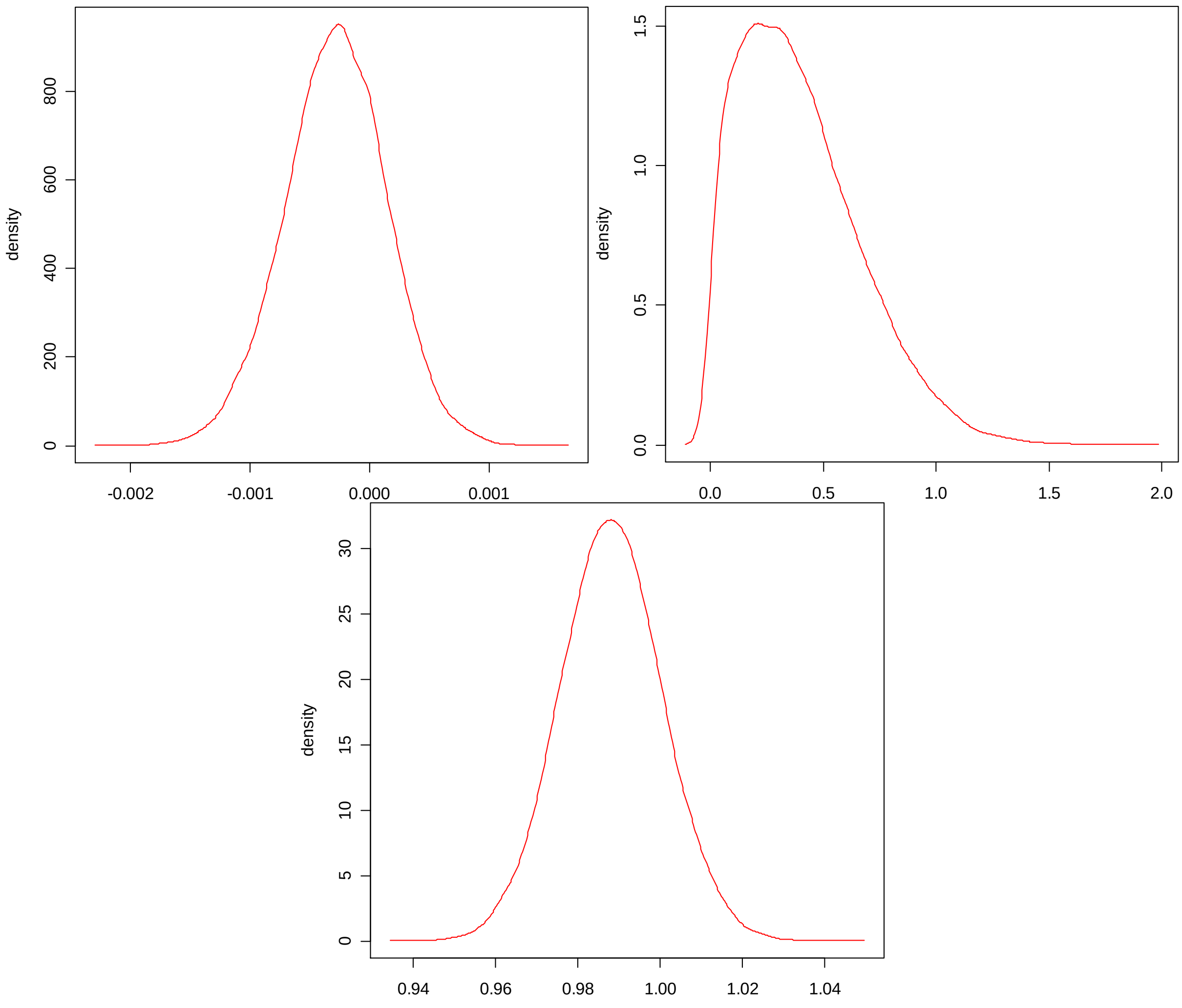

Fig. 8. Parameters $\alpha$ , $\beta$ and $\psi_{(6,3)}$ for the room A

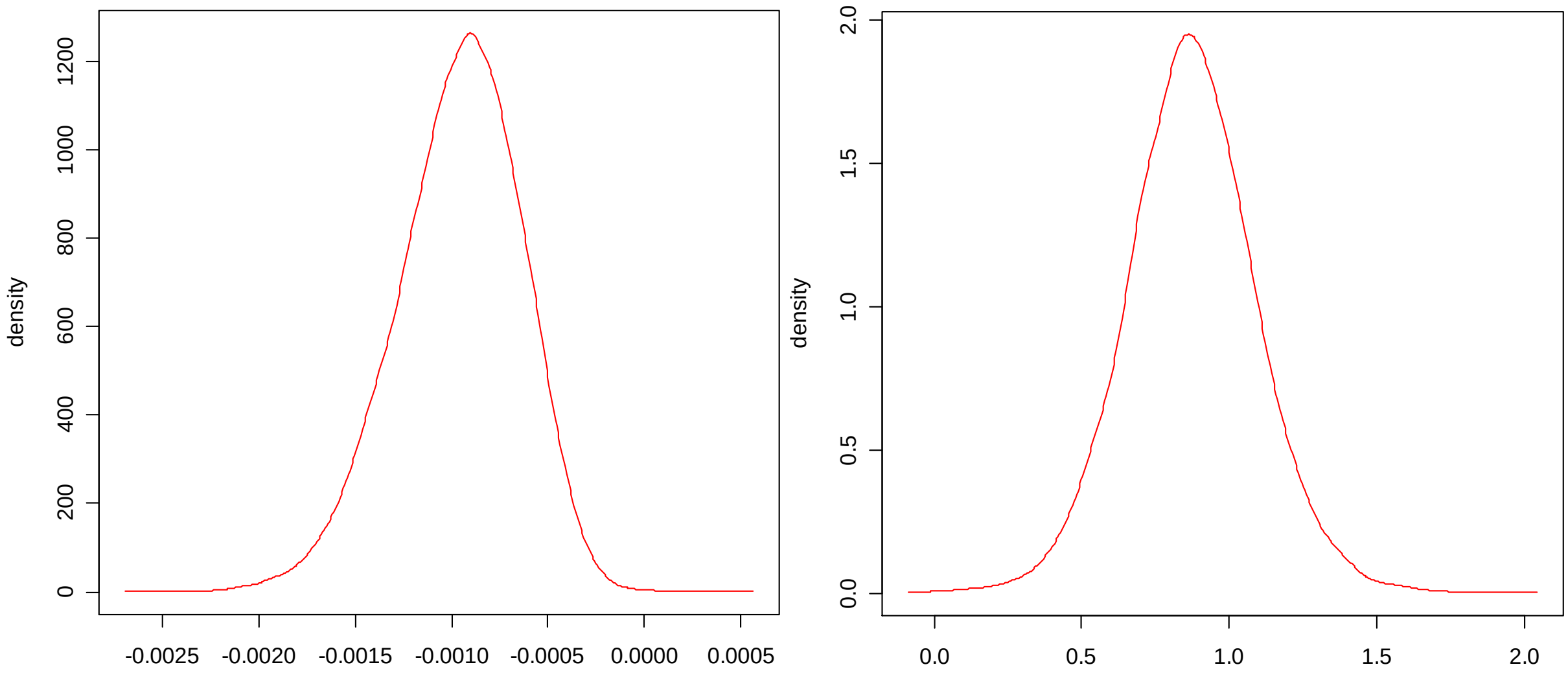

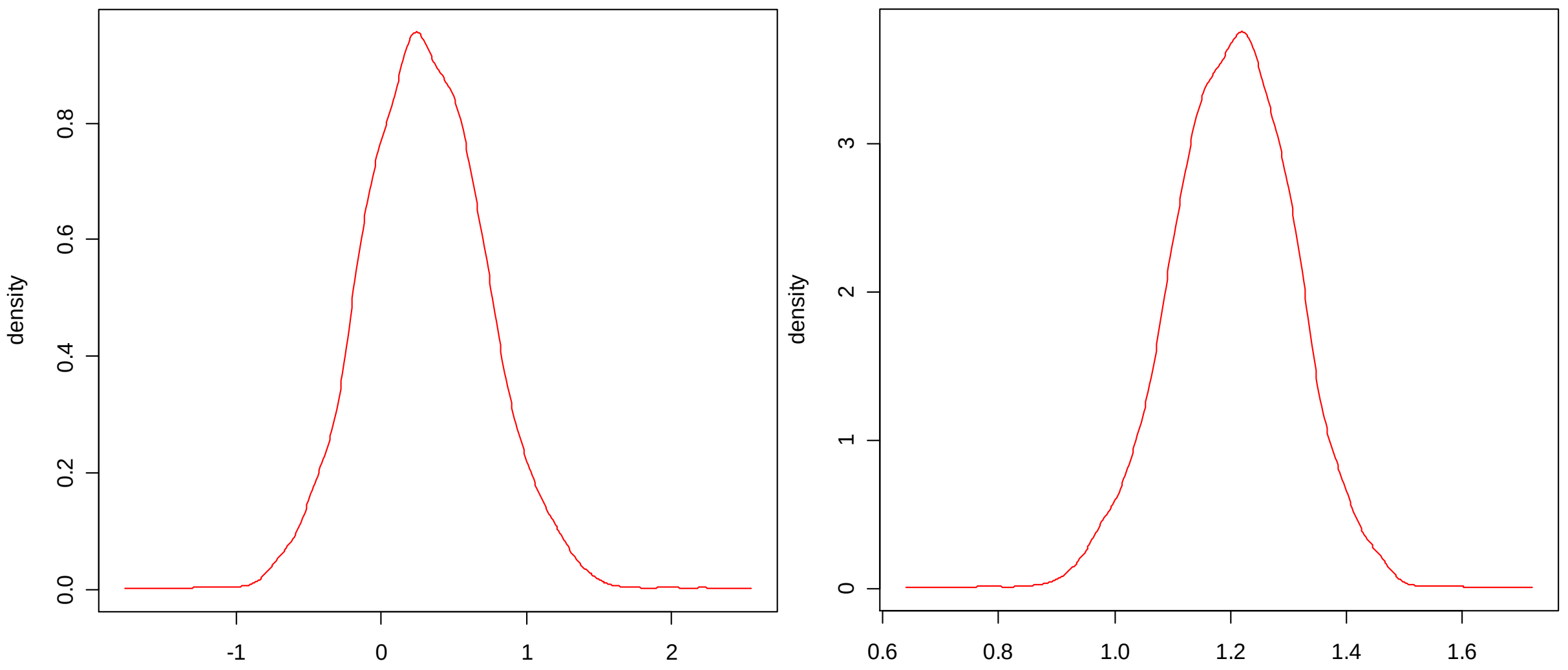

Fig. 9. Parameters $\alpha$, $\beta$, $\psi_{(5,1)}$ and $\psi_{(7,4)}$ for the room B

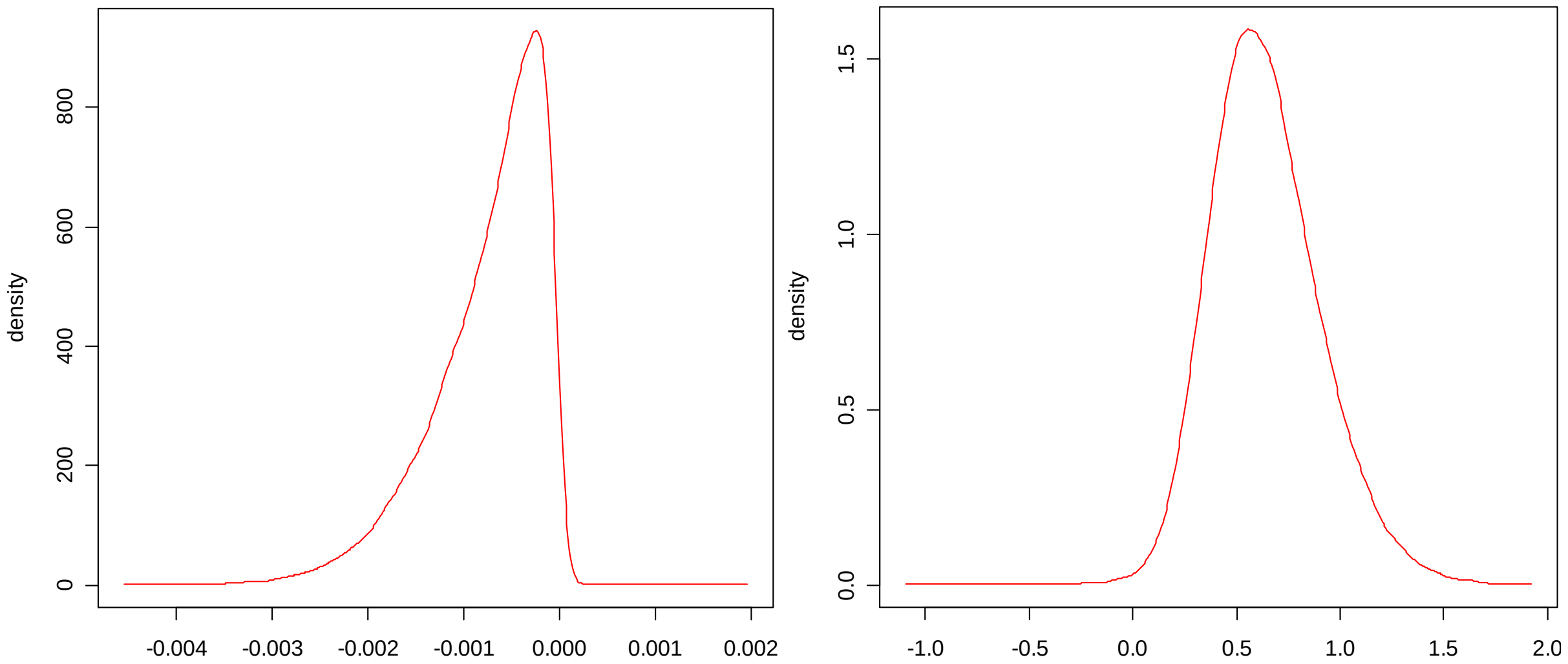

Fig. 10. Parameters $\alpha$ and $\beta$ for the room C

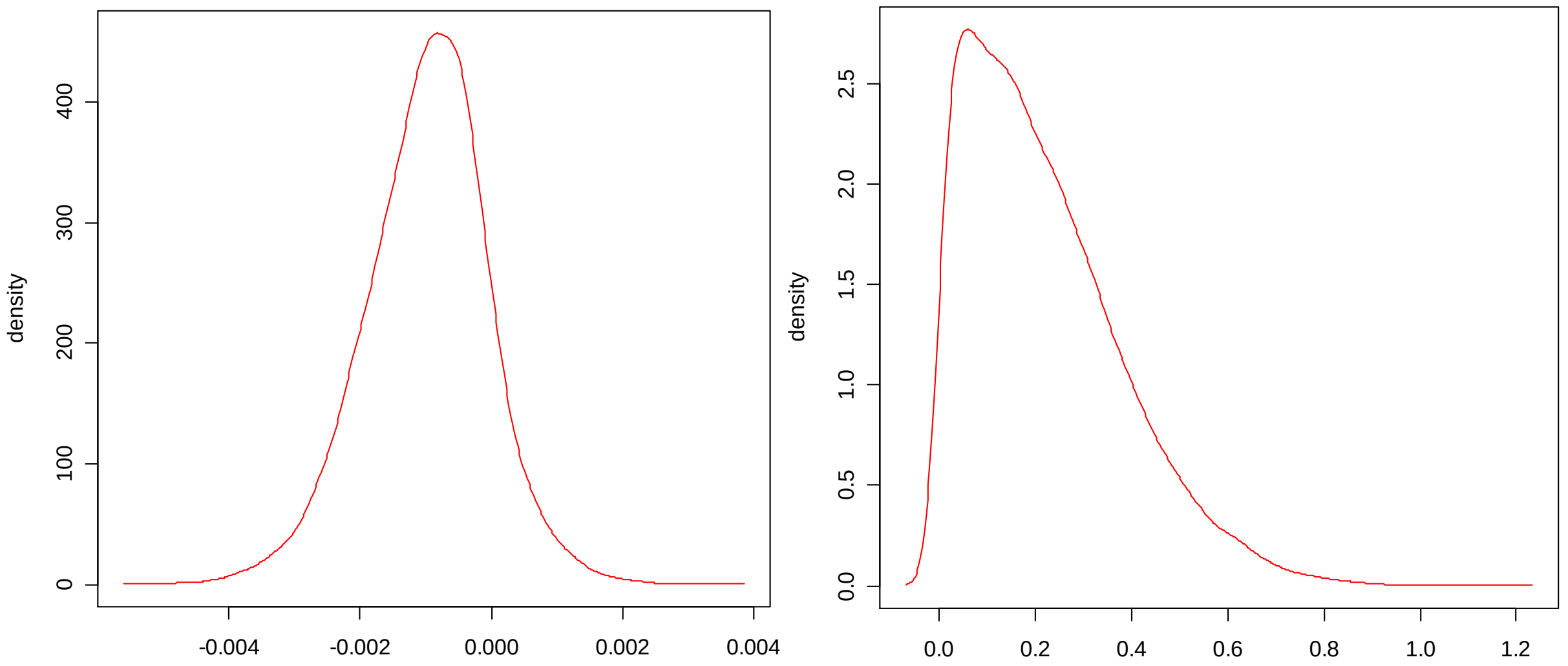

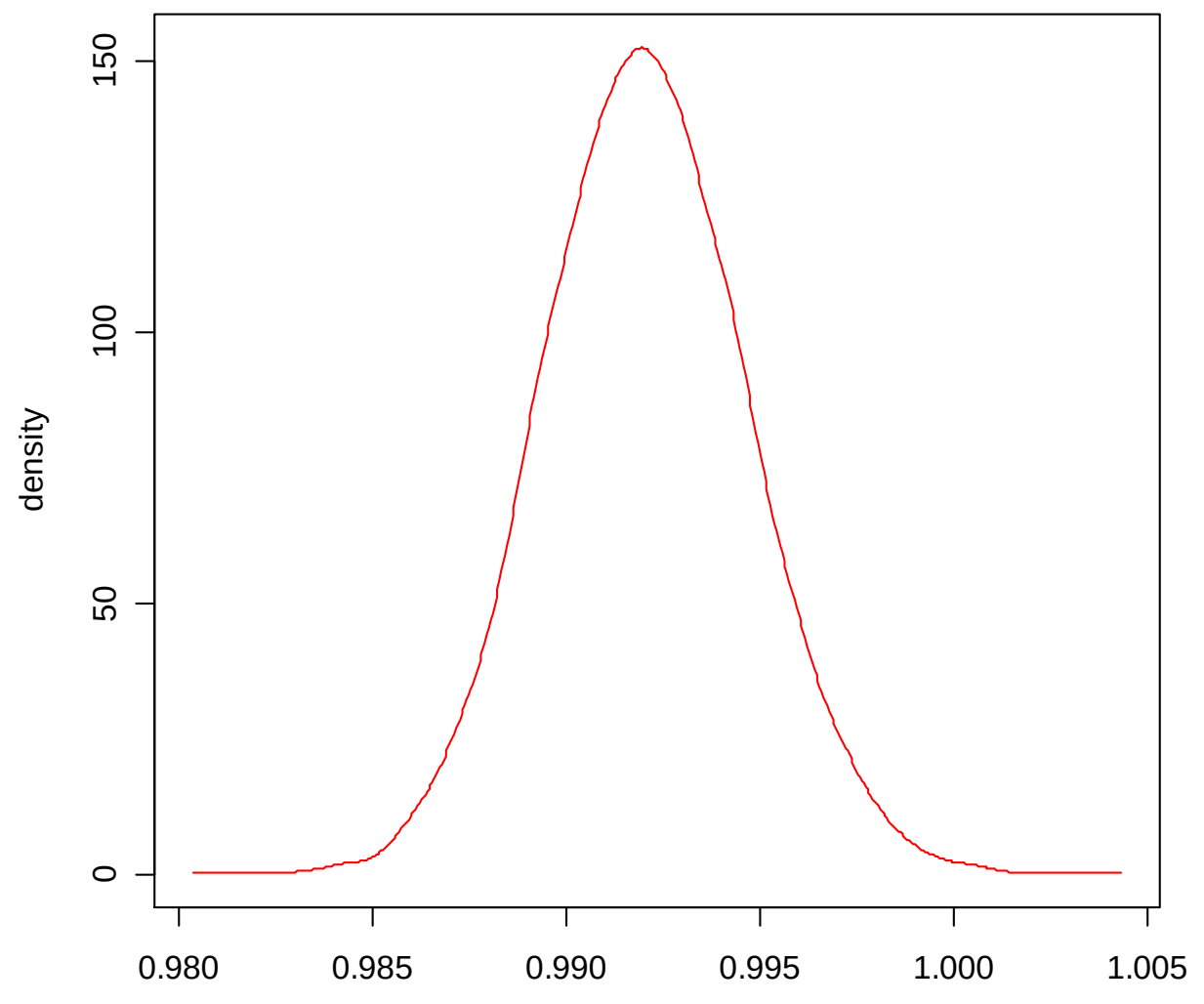

Fig. 11. Parameters $\alpha$ , $\beta$ and $\psi_{(8,3)}$ for the room D

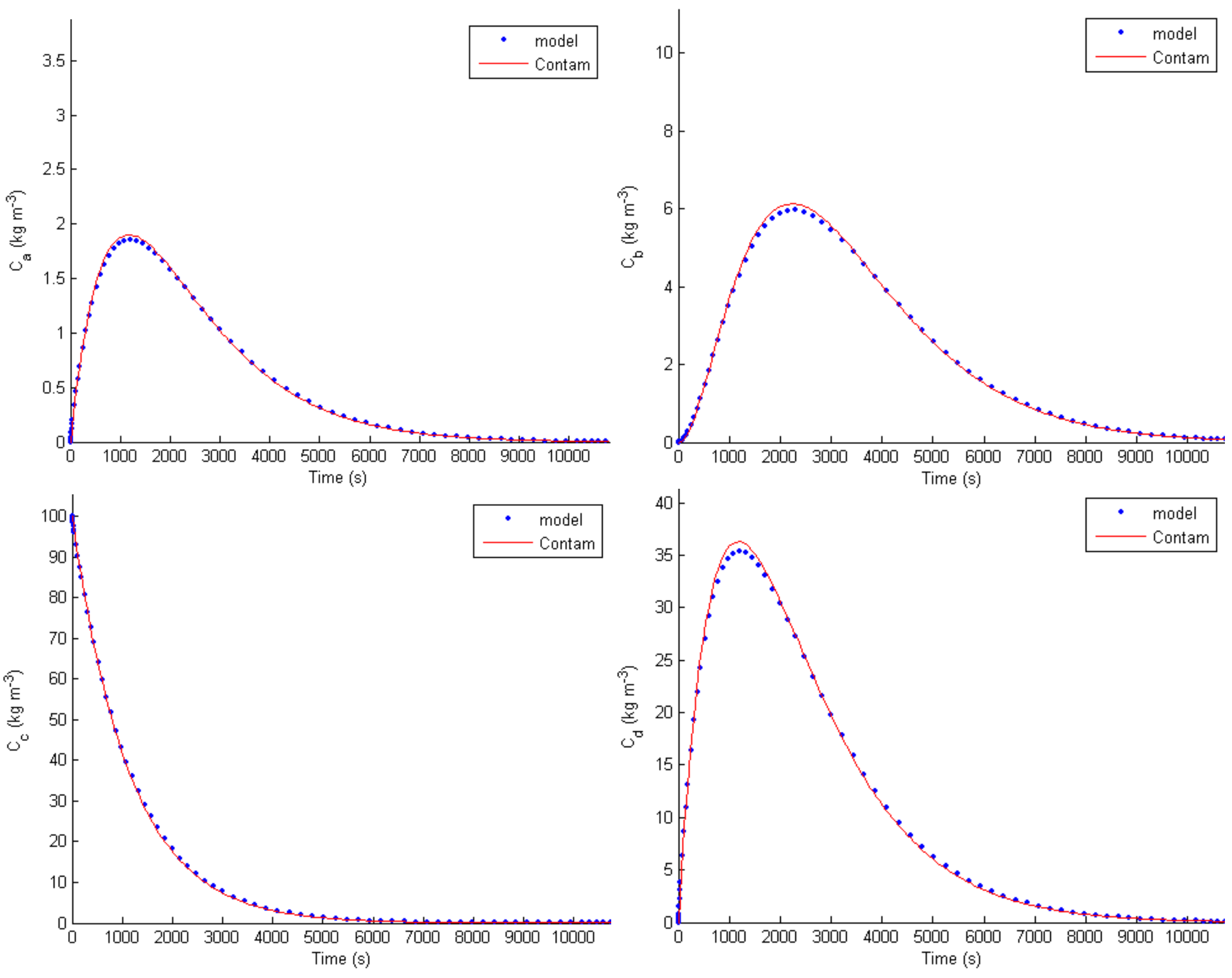

Fig. 12. Comparison of concentration results for each rooms

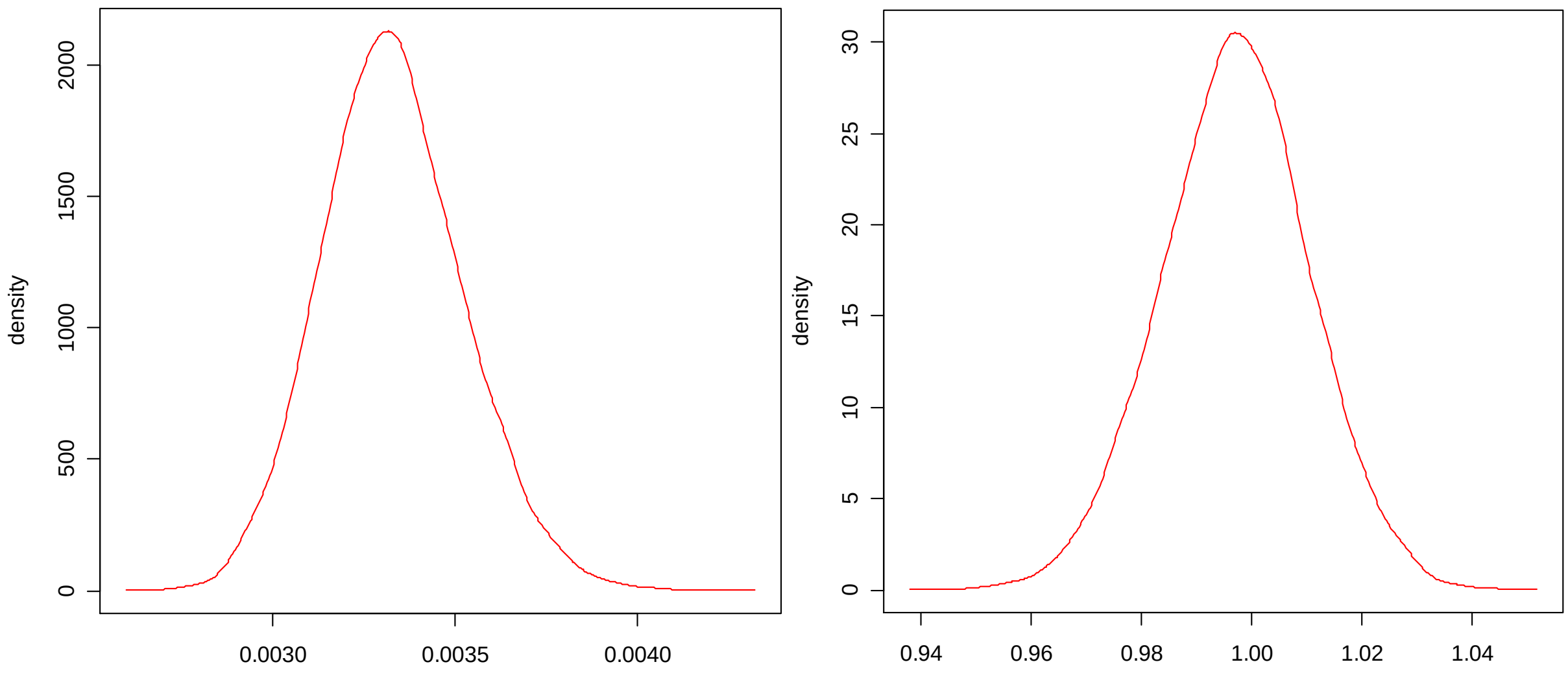

Fig. 13. Parameters $\alpha$ , $\beta$ for the room A

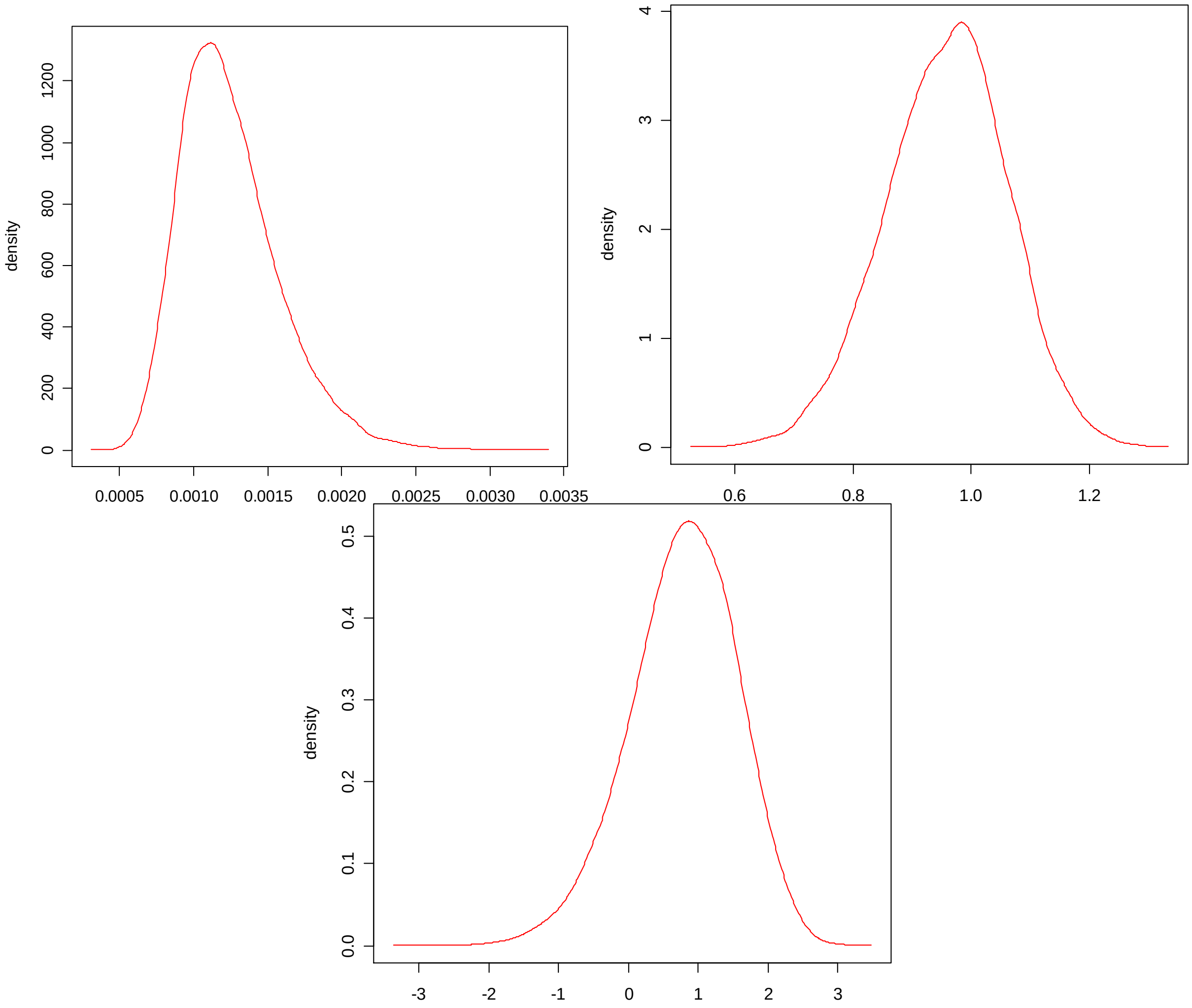

Fig. 14. Parameters $\alpha$ , $\beta$ and $\psi_{(9,1)}$ for the room B

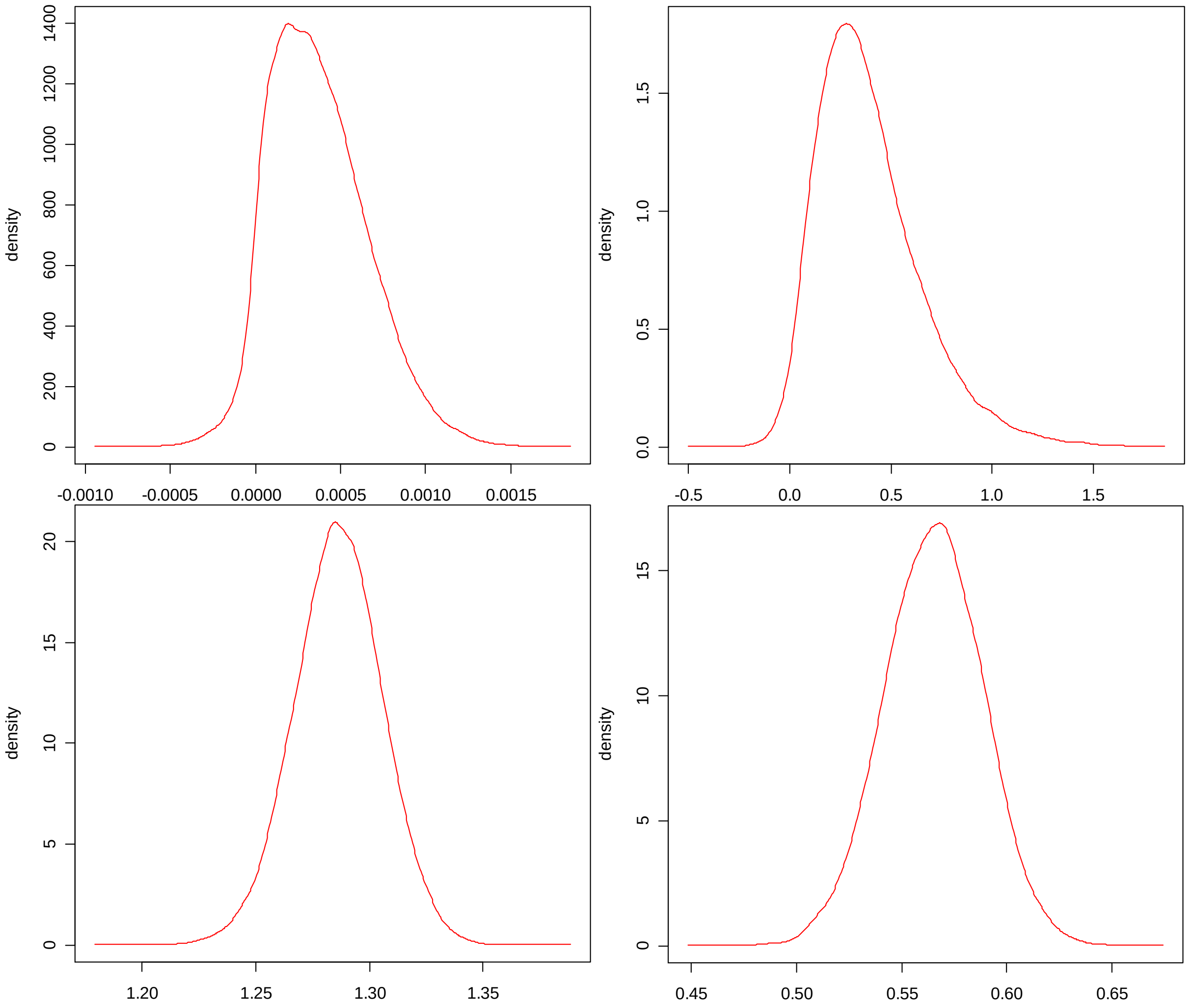

Fig. 15. Parameters $\alpha$ , $\beta$ , $\psi_{(10,2)}$ and $\psi_{(8,4)}$ for the room C

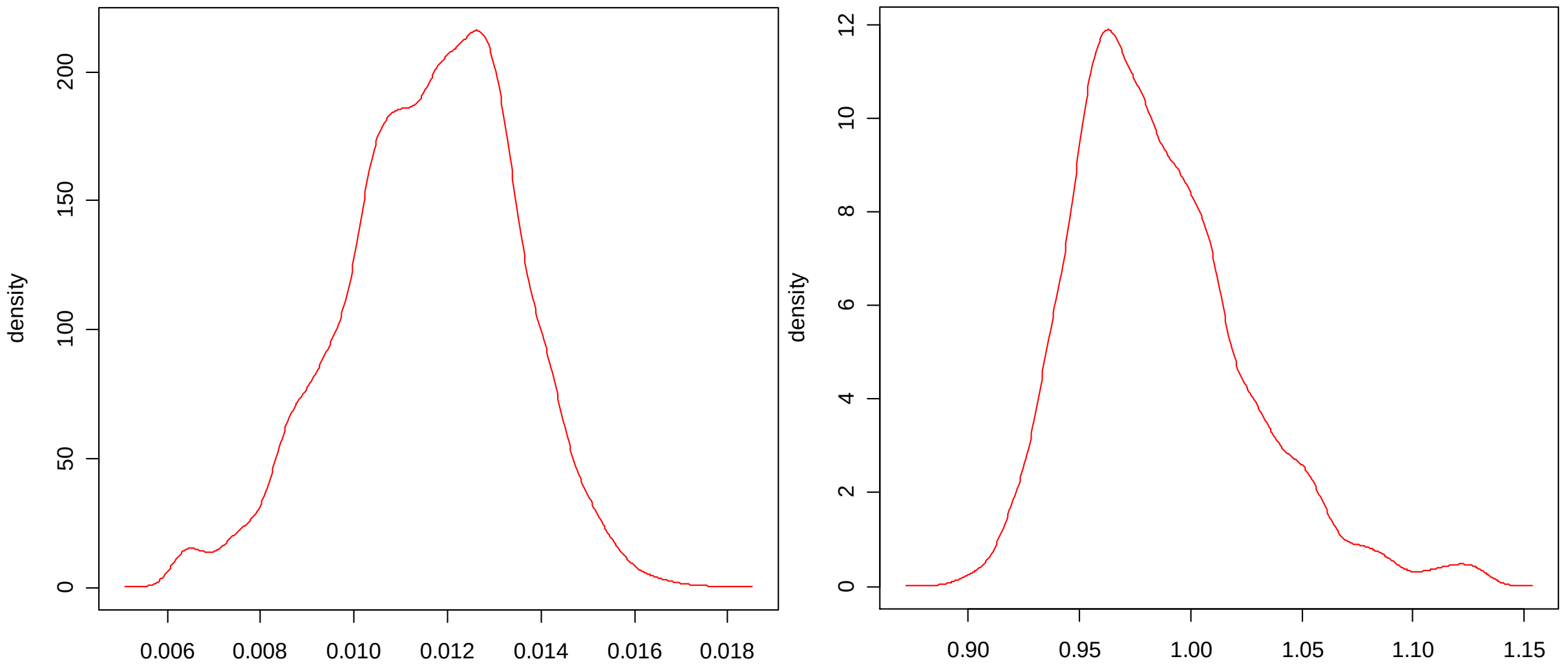

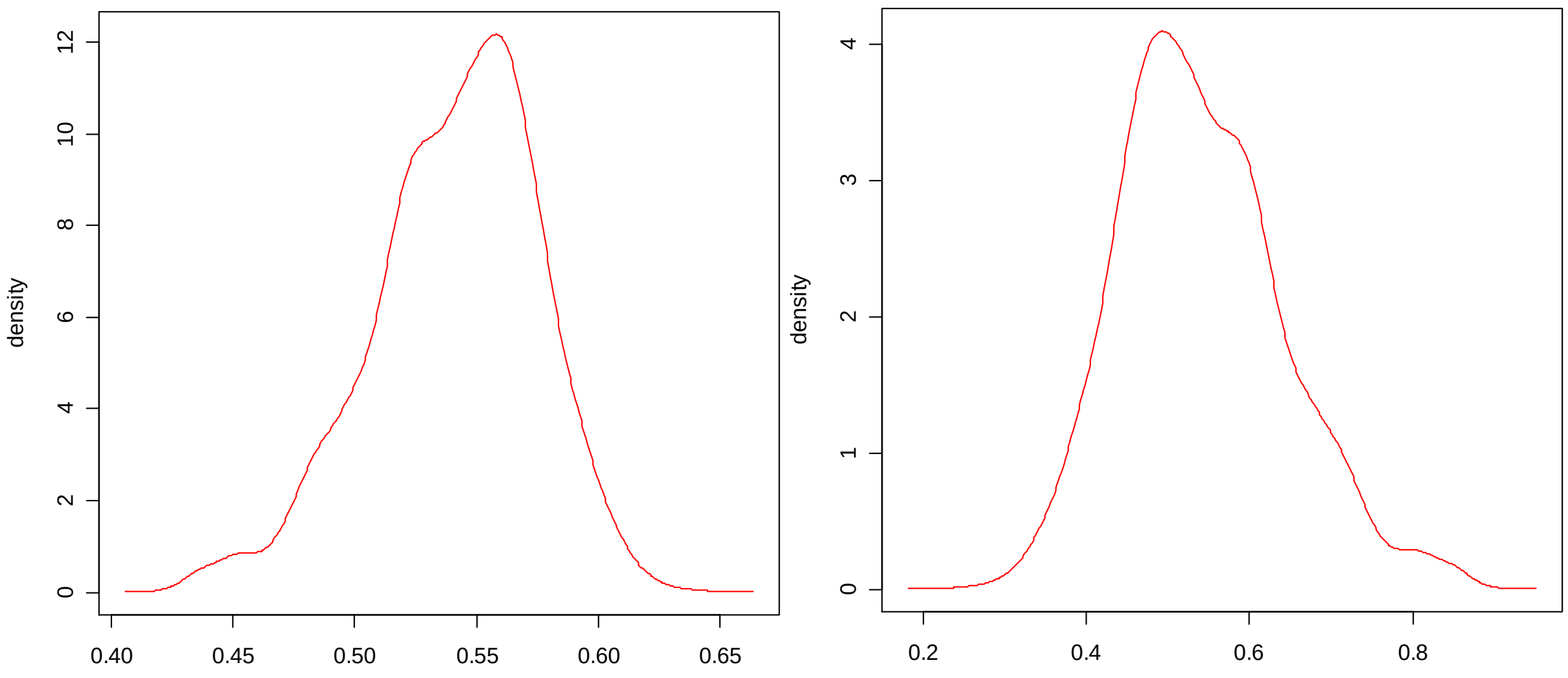

Fig. 16. Parameters $\alpha$ , $\beta$ , $\psi_{(6,1)}$ and $\psi_{(7,2)}$ for the room D

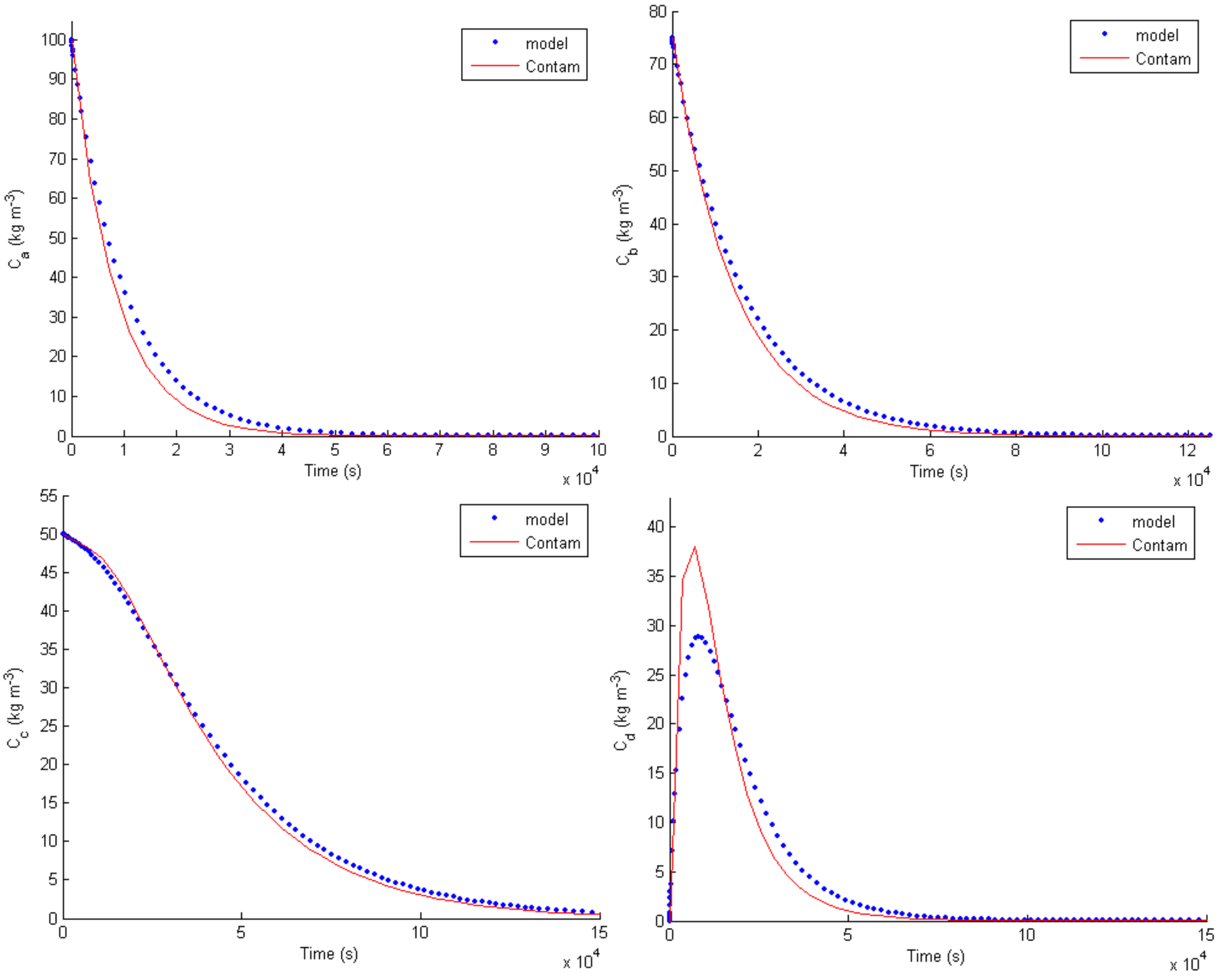

Fig. 17. Comparison of concentration results for each room

Je me demande s'il faut mettre ce système d'équations avec les paramètres ou pas (à la place de l'autre)

Le modèle IEA:

$$V_1 \frac{\partial C_1}{\partial t} = F_{1,0} m_1 C_0 - m_6 C_1 - m_9 C_1 + S_1$$
$$V_2 \frac{\partial C_2}{\partial t} = F_{2,0} m_2 C_0 + F_{9,1} m_9 C_1 - m_{10} C_2 - m_7 C_2 + S_2$$
$$V_3 \frac{\partial C_3}{\partial t} = F_{10,2} m_{10} C_2 + F_{8,4} m_8 C_4 - m_3 C_3 + S_3$$
$$V_4 \frac{\partial C_4}{\partial t} = F_{6,1} m_6 C_1 + F_{7,2} m_7 C_2 + F_{5,0} m_5 C_0 - m_8 C_4 - m_4 C_4 + S_4$$

Le modèle RAO:

$$V_1 \frac{\partial C_1}{\partial t} = m_2 C_0 + F_{63} m_6 C_3 - m_5 C_2 + \alpha C_1^{\beta} + S_1$$
$$V_2 \frac{\partial C_2}{\partial t} = F_{51} m_5 C_1 + F_{74} m_7 C_4 - m_3 C_2 + \alpha C_2^{\beta} + S_2$$
$$V_3 \frac{\partial C_3}{\partial t} = F_{10} m_1 C_0 - m_6 C_3 - m_8 C_3 + \alpha C_3^{\beta} + S_3$$
$$V_4 \frac{\partial C_4}{\partial t} = F_{83} m_8 C_3 - m_4 C_4 - m_7 C_4 + \alpha C_4^{\beta} + S_4$$